\documentclass[a4paper,10pt,fleqn]{article}
\pdfoutput=1 % if your are submitting a pdflatex (i.e. if you have
\usepackage{jheppub} % for details on the use of the package, please
\usepackage[utf8]{inputenc}
\usepackage{amsmath}
\def\gsim{\mathrel{\rlap{\lower4pt\hbox{\hskip1pt$\sim$}}
    \raise1pt\hbox{$>$}}}         %greater than or approx. symbol
\def\lsim{\mathrel{\rlap{\lower4pt\hbox{\hskip1pt$\sim$}}
    \raise1pt\hbox{$<$}}}         %less than or approx. symbol
\newcommand{\ptcut}{p_t^{\text{cut}}}
\newcommand{\pt}[1]{p_{t#1}}
\newcommand{\ptt}{p_{t}}

\newcommand{\kt}[1]{k_{t{#1}}}

\newcommand{\pth}{\tilde{p}_t}
\newcommand{\Rmax}{R_{\text{max}}}

\newcommand{\p}{i}

\newcommand{\sigmatilde}{\tilde{\sigma}}

\usepackage{multirow}
\usepackage{pifont}
\newcommand{\xmark}{\ding{55}}

\usepackage{geometry}
\title{\boldmath The single-jet inclusive cross-section and its definition}

\author[a,b]{Matteo Cacciari}
\author[c]{, Stefano Forte}
\author[d]{, Davide Napoletano}
\author[d]{, Gregory Soyez}
\author[a,b,c]{ and Giovanni Stagnitto}

\affiliation[a]{Universit\'e Paris Diderot, F-75013 Paris, France}
\affiliation[b]{Sorbonne Universit\'e, CNRS, Laboratoire de Physique Th\'orique et Hautes \'Energies, LPTHE, F-75005 Paris, France}
\affiliation[c]{Tif Lab, Dipartimento di Fisica, Universit\`a di Milano and INFN, Sezione di Milano, Via Celoria 16, I-20133 Milano, Italy.}
\affiliation[d]{IPhT, CEA Saclay, CNRS, Universit\'e Paris-Saclay, F-91191 Gif-sur-Yvette cedex, France.}

\emailAdd{cacciari@lpthe.jussieu.fr}
\emailAdd{forte@mi.infn.it}
\emailAdd{davide.napoletano@ipht.fr}
\emailAdd{gregory.soyez@ipht.fr}
\emailAdd{giovanni.stagnitto@lpthe.jussieu.fr}

\abstract{We investigate some well-known problematic aspects of the single-jet
  inclusive cross-section, specifically its non-unitarity and the
  possibly related issue of  apparent
  perturbative instability at low orders. We study and clarify their
  origin by introducing possible alternative weighted definitions of the
  observable which restore unitarity. We show that
  the perturbative instability of the standard definition is
  an accidental artefact of the smallness of the NLO $K$ factor which
  only manifests itself for values of the jet radius in the range
  $R\sim 0.3-0.6$, and that its non-unitarity is necessary in order to ensure
  cancellation of logs of the momentum cutoff used in the jet definition. We
  also show that alternative unitary definitions do not
  have better perturbative properties compared to the conventional non-unitary
  definition, while suffering from lack of cancellation of large logs.}

\begin{document}
\begin{flushright}
  TIF-UNIMI-2019-8
\end{flushright}
\maketitle

\section{Introduction}\label{sec:intro}
The single-jet inclusive cross-section has been
used for now over thirty years~\cite{Martin:1987vw} for the
determination of parton distributions. As an observable, it is defined in a
deceptively simple way~\cite{Aversa:1988fv,Ellis:1988hv}:
count all jets which fall in any given kinematic bin and add them up.
While this definition is remarkably simple, a minutes' reflection
shows that it has a somewhat peculiar and perhaps undesirable feature. Namely,
it is not unitary: each event is counted
more than once, so that the integral of the differential
cross-section does not yield the total
cross-section. The recent computation of the next-to-next-to-leading
order (NNLO) corrections to this observable~\cite{Currie:2016bfm,Currie:2018xkj} has shown
another seemingly problematic aspect: the scale dependence of
the result is not significantly reduced and the size of the $K$ factor
does not significantly decrease when going from NLO to NNLO,
at least with certain scale choices, which suggests a possible perturbative
instability.

In Ref.~\cite{Currie:2018xkj} the perturbative properties
of this observable were extensively studied, in particular by
a numerical analysis of the contributions to individual jet bins with
a variety of computational setups (such as the choice of scale and of
jet radius). Here we approach the problem of understanding the
behavior of this observable from a somewhat different
point of view: namely, by trying to see how it behaves upon changes of
its definition, specifically motivated by an attempt to correct for
its non-unitarity. We then study the properties of this family of new,
unitary definitions both numerically, and analytically in a simple
collinear approximation.
Our analysis focuses on the general properties of the observable,
of which we strive to understand the main qualitative
features. We thus base our discussion on NLO calculations,
whose structure is easier to handle both from a numerical and an
analytic point of view, though we aim at understanding their
general properties at any perturbative order. An explicit study of
NNLO results (which are not publicly available anyway) such as
already presented in  Ref.~\cite{Currie:2018xkj}, as needed for
state-of-the-art precision phenomenology, is outside our
scope and goals. Nevertheless, we will comment when needed on the
validity of our results at higher orders, and we 
have explicitly checked their robustness in several cases  at  NNLO,
which we have been 
able to obtain from a NLO code by calculating differences in which
missing double-virtual contributions cancel.

Our main conclusion is that what seems to be an undesirable feature,
namely the non-unitarity of the standard definition, automatically
guarantees that results are stable upon changes of
the cutoff momentum scale used to in order to define a jet, i.e.\ the
minimum momentum that a jet must carry. Introducing an
alternative, unitary definition of the cross-section, 
preserving insensitivity to the momentum cutoff, is nontrivial, and
requires that
unitarity be made compatible with independence of  
the number of jets: we will show two examples demonstrating how this could
 be achieved.

 On the other hand,
what may appear to be a lack of perturbative
convergence when going from NLO to NNLO, with the NNLO
correction~\cite{Currie:2018xkj} larger or of the same order of the
NLO one, 
is actually a manifestation of
the fact that the NLO correction of the cross-section depends on $R$
in such a way that it changes sign around $R\sim0.4$, and it is thus
accidentally small, with small theoretical uncertainties, around
$R=0.4$.
The perturbative properties of alternative, weighted definitions are
generally similar to that of the standard definition, though
often worse, for reasons closely related to the sensitivity to
the transverse momentum cutoff.

The outline of this paper is the following. First, in
Sect.~\ref{sec:definition} we discuss the standard definition of the
cross-section and its non-unitarity, and present a family of
alternative, unitary definitions. Then, in Sect.~\ref{sec:numerics} we
compare results obtained using various definitions at NLO.
In Sect.~\ref{sec:analytics} we show how the results of the
previous section can be understood in terms of an analytical calculation.
Finally, we draw our conclusions in Sect.~\ref{sec:conclusion}.

\section{The single-jet inclusive cross-section and its definition}
\label{sec:definition}

The single-jet inclusive cross-section is defined in terms of the
differential cross-section
$\frac{d \sigma_{N\,\text{jets}}}{d \pt{1}\dots d \pt{N}}$ for producing
$N$ jets (after cuts) with transverse momenta $\pt{i}$, as
\begin{align}
  \label{eq:stdef0}
  \frac{d \sigma}{d \ptt} &= \sum_N \frac{d \sigma_{N\,\text{jets}}}{d  \ptt}\\
  \label{eq:stdef}
  \frac{d \sigma_{N\,\text{jets}}}{d
    \ptt} &=  \int d \pt1 \dots d \pt{i} \dots d \pt{N}\,
  \frac{d \sigma_{N\,\text{jets}}}{d \pt{1} \dots d \pt{i} \dots d
    \pt{N}} F_N\left[\pt{1},\dots,\pt{N};\ptt\right] ,
\end{align}
where $F_N$, for a standard definition, is given by 
\begin{equation}\label{eq:binning}
F_N^{\text{std}}\left[\pt{1},\dots,\pt{N};\ptt\right] =\sum_{i=1}^N  \delta(\pt{i}-\ptt),
\end{equation}
and it fills the bin with transverse momentum $p_t$ by picking all
contributions from the fully differential $N$-jets cross-section.
The sum in Eq.~(\ref{eq:stdef0}) runs over the number of jets in each
event that pass some kinematic cut.
The sum over the total number of jets starts with $N=1$ (the $N=0$
case gives of course no contribution) and goes up to two at leading order (LO),
three at NLO, and generally $p+2$ at N$^p$LO.

It is clear that the inclusive-jet cross-section
defined in this way is not unitary, in that its integral over $\ptt$ does not give
the total number of scattering events per unit flux per unit
time within a given fiducial region.  Indeed, with this definition, when filling a
histogram in $\ptt$, an event with $N$ jets is binned $N$ times.
This lack of unitarity may be cause of concern: one is used to
the fact that the unitarity
of the total partonic cross-section is crucial in order to ensure its infrared
finiteness, given that infrared singularities cancel between terms
with different numbers of final--state partons. On the other hand,
infrared finiteness of the $N$-jet cross-section is ensured by the use
of a jet definition, so the question is
really whether this definition leads to a good perturbative behavior.

In order to address the question in a quantitative way,
we generalize the definition of the single-jet inclusive cross-section by
introducing jet weights that render the cross-section unitary.
Namely, we modify the definition Eq.~(\ref{eq:stdef}) by introducing
weights in the definition of the 
function $F_N$, Eq.~(\ref{eq:binning}):
\begin{equation}\label{eq:stw}
F_N\left[\pt{1},\dots,\pt{N};\ptt\right] =\sum_{i=1}^N
\delta(\pt{i}-\ptt) w^{(N)}(\ptt;\pt{1},\dots,\pt{N})
\end{equation}
The choice $w^{(N)} = 1$ represents the standard non unitary
definition Eq.~(\ref{eq:binning}). The choice $w^{(N)} = 1/N$ restores
unitarity,
but has undesirable discontinuities whenever the kinematics of
the final state changes in such a way that the number of jets
jumps from $N$ to $N+1$.
In this work, we consider a set of weights defined as
\begin{equation}
  \label{eq:wexpr}
  w^{(N)}(\ptt;\pt{1},\dots,\pt{N}) = \left\{
    \begin{split}
      & 1 \quad & ({\text{standard}}) \\
      & \frac{\ptt^r}{\sum_{j=1}^N\pt{j}^r} \quad & ({\text{weighted}})
    \end{split}
  \right.
\end{equation}
where  $\pt{j}$ is the transverse momentum of the $j$-th jet.
All weighted choices lead to a unitary definition.

We consider specifically three families of definitions of these
weights, according to which jets are included when constructing the weights.
\begin{itemize}
\item {\it A: jets above }$\ptcut$\\
  Only jets  with $\ptt\geq\ptcut$ are included in the definitions of
  $F_N$ Eq.~(\ref{eq:stw}). In particular, this implies that the sum
  in the denominator of Eq.~(\ref{eq:wexpr}) includes only 
  jets for which $\pt{j} \geq \ptcut$. When $r=0$ this reduces to the
  simplest unitary choice with all weights equal to $1/N$.
\item {\it B: all jets}\\ $F_N$ includes all the jets but the
  numerator in the weight definition, Eq.~(\ref{eq:wexpr}), only
  includes jets above $\ptcut$. In particular, the denominator in
  Eq.~(\ref{eq:wexpr}) sums over all jets.
  This definition is infrared safe only for $r>0$.  While this
  definition may seem unphysical, in practice it corresponds to having
  a $\ptcut$ that is small compared to the $\ptt$ value of the first
  bin one is interested in.
\item {\it  C: two leading jets}\\ Only the first two leading jets in
  $\ptt$
  % jets (i.e. the jets with largest $p_t$)
  are included in the definition of both $F_N$ and the
  weights,
  so  $N=2$ in both Eqs.~(\ref{eq:stw}) and~(\ref{eq:wexpr}).
  In this case  we
  consider the two leading jets independently on whether their $\ptt$
  is larger or smaller than a possible $\ptcut$.
\end{itemize}

These definitions are ``unitary'' in the sense that the 
weights add up to one.
This implies that, with the first definition, integrating over $p_t$
gives the total cross-section to
have at least one jet above $\ptcut$.
For the second definition (with $\ptcut\to 0$ or an explicit underflow
bin) and for the third
definition, one instead gets the total $pp$ cross-section.
To keep the discussion simple, we do not impose any rapidity cut in the studies
carried on in this paper. Nevertheless each of the previous
definitions could be extended to the case in which a rapidity cut is
introduced.
Note that in the case of the
  third definition, a rapidity cut could change what the leading
  jets are. To avoid potential issues, in particular for $r<0$ which
  is more sensitive to small $p_t$, one might have in practice to
  impose an additional dijet selection cut (similar to what is already
  done when studying e.g. the dijet invariant mass).

To highlight the various features we are interested in studying in
this work, it is useful to consider different
ways of organizing the perturbative calculation of the single-jet
inclusive cross-section at N$^p$LO accuracy.
This can, in fact, be written as a sum of contributions, each of
order $\alpha_s^{2+k}$, $k=0,\dots,p$,
% and up to $\alpha_s^{2+p}$,
assuming that the leading-order (LO) process is of order $\alpha_s^2$:
\begin{equation} \label{eq:sigmaNKLO}
  \frac{d \sigma^{\text{N}^p\text{LO}}}{d \ptt} =
  \sum_{k=0}^{p}  \frac{d \sigma^{(k)}}{d \ptt}.
\end{equation}
Furthermore, it is useful to think about the order $\alpha_s^{k+2}$
contribution in two different ways.
%
% Furthermore each of the order $k$ contributions can be thought of in
% two different ways.
%
The first is as a sum of contributions with a different number
of jets, as we have done in Eq.~(\ref{eq:stdef}). In such a case, the
$k$-th order contribution to the cross-section is built out of
terms containing at most $k+2$ jets i.e.\ two at LO ($k=0$),
three  at NLO ($k=1$) and so forth:
\begin{equation}\label{eq:sigmaknNjet}
  \frac{d \sigma^{(k)}}{d \ptt}
  = \sum_{N=1}^{k+2}
  \frac{d \sigma^{(k)}_{N\,\text{jets}}}{d \ptt}.
\end{equation}
Eq.~(\ref{eq:sigmaknNjet}) is the same as Eq.~(\ref{eq:stdef0}),
  but for the $k$-th order contribution only.
However, in order to understand the perturbative behavior of the cross-section it
also useful to break it up into the contribution from the
jet with the largest $\ptt$ (leading, or first jet), the jet with the
second largest $\ptt$ (subleading, or second jet), and so on:
\begin{equation}\label{eq:sigmaknN}
  \frac{d \sigma^{(k)}}{d \ptt}
  = \sum_{n=1}^{k+2}
  \frac{d \sigma^{(k)}_{n\text{-th\,jet}}}{d \ptt}.
\end{equation}
In Eq.~(\ref{eq:sigmaknNjet}), $d \sigma^{(k)}_{N\,\text{jets}}/d \ptt$ is
the contribution to the cross-section coming from configurations with $N$ jets,
while in Eq.~(\ref{eq:sigmaknN}) $d \sigma^{(k)}_{n\text{-th jet}}/d \ptt$
is the contribution coming from the $n$-th leading jet. 
The range of the sum is the same in both cases and it is equal to the maximum
number of jets that can be produced at a given perturbative order $k$.

\section{Comparing definitions of the cross-section}
\label{sec:numerics}

In order to study the effects of the various unitary
definitions, Eq.~\eqref{eq:wexpr}, we start by simply comparing results
obtained in each case for the NLO $K$ factors and individual jet contributions. 
This way, we can see how
imposing unitarity affects the $\ptt$ distribution of the single-jet
inclusive cross-section. In  
Section~\ref{sec:analytics} we then turn to analytic arguments,
both in general and in  a collinear approximation. While the
discussion presented here is mostly at NLO, we have explicitly
checked that our results persist through NNLO, by computing at NNLO
the difference of the cross-section with the various definitions
that we consider, which
can be done using public NLO codes.

All results presented in this section are obtained using the following
setup.
Computations up to NLO are performed using
\texttt{NLOJET++}(v4.1.3)~\cite{Nagy:2003tz,Nagy:2001fj} for  $pp$ collisions,
with center of mass energy $\sqrt{s}=13$~TeV.
Parton distribution functions are taken from the
NNPDF3.1~\cite{Ball:2017nwa} set at NNLO, with $\alpha_s(M_Z)=0.118$, and
interfaced using the \texttt{LHAPDF} library (v6.1.6)~\cite{Buckley:2014ana}.
Jets are clustered using the anti-$k_t$ algorithm~\cite{Cacciari:2008gp},
as implemented in \texttt{FastJet}(v3.3.2)~\cite{Cacciari:2011ma}, with
$R = 0.4$, unless otherwise specified.

The dependence on the choice of central factorization and
renormalization scale  (see e.g.\ the discussion
in~\cite{Currie:2018xkj}) is studied by considering  three options:
(i) the {\it average dijet scale},
\begin{equation}
\ptt^\text{(avg)} = \frac{\pt{1}^{(R=1)}+\pt{2}^{(R=1)}}{2},\label{eq:avscale}
\end{equation}
where $\pt{1,2}^{(R=1)}$ are the transverse momenta of the two
leading jets clustered with a radius $R=1$~\cite{Dasgupta:2016bnd},
(ii) the {\it partonic scalar $k_t$ halved},
\begin{equation}
\frac{\hat{H}_T}{2} = \frac{1}{2} 
    \sum_{i=1}^{n\text{-partons}}  \kt{i}, \label{eq:partscale}
\end{equation}
suggested as an
optimal scale choice in \cite{Currie:2018xkj},
and (iii) the {\it leading jet $p_t$}, $\ptt^\text{(max)}$, defined
as $\ptt$ of the leading $R=1$ jet.
% the largest $\ptt$ obtained with an $R=1$ jet clustering.

For each  choice of central scale, uncertainty bands are
obtained with the 7-point scale variation rule~\cite{Cacciari:2003fi}.
As  noted in~\cite{Dasgupta:2016bnd} and as we discuss below,
the uncertainty bands around the NLO prediction are unnaturally small
because of an unphysical cancellations in scale dependence between
the production of hard partons, a large angle process,
and their fragmentation into jets, a small angle one.
A more reliable estimate could be obtained by
factoring the cross-section for producing a small-radius jet
into the cross-section for the initial partonic scattering
and the fragmentation of the parton to a jet,
considering separately the uncertainties of these two processes
and summing them in quadrature.
This option has been studied in~\cite{Dasgupta:2016bnd} and in great
details more recently in~\cite{Bellm:2019yyh}.
In this paper, we have checked the effects of decorrelated scale variation
on the weighted definitions, and we briefly comment on this below.
% 
% This option has been recently studied in great detail for the standard
% definition in 
% Ref.~\cite{Bellm:2019yyh}, to which we refer for more details.
% We
% have checked  the effects of decorrelated scale variation
% on the weighted definitions, and
% we will briefly comment on this below.

\subsection{Standard (non-unitary) definition}\label{sec:numerics-std}
We start by discussing some well-known results for the
standard definition. As mentioned, we focus  on two
observables: the total NLO $K$ factor, and the individual $n\text{-th}$-leading
jet NLO $K$ factor as a function of $\ptt$,
\begin{equation}
  K = \sum_{n=1}^3 K_{n}\,, \quad \text{with}
  \quad K_{n} = \frac{{\rm d}\sigma^{\text{NLO}}_{n\text{-th\,jet}}}{{\rm d}\sigma^{\text{LO}}}\,.
\end{equation}

\begin{figure}[t]
  \centering
  \includegraphics[width=0.45\textwidth]{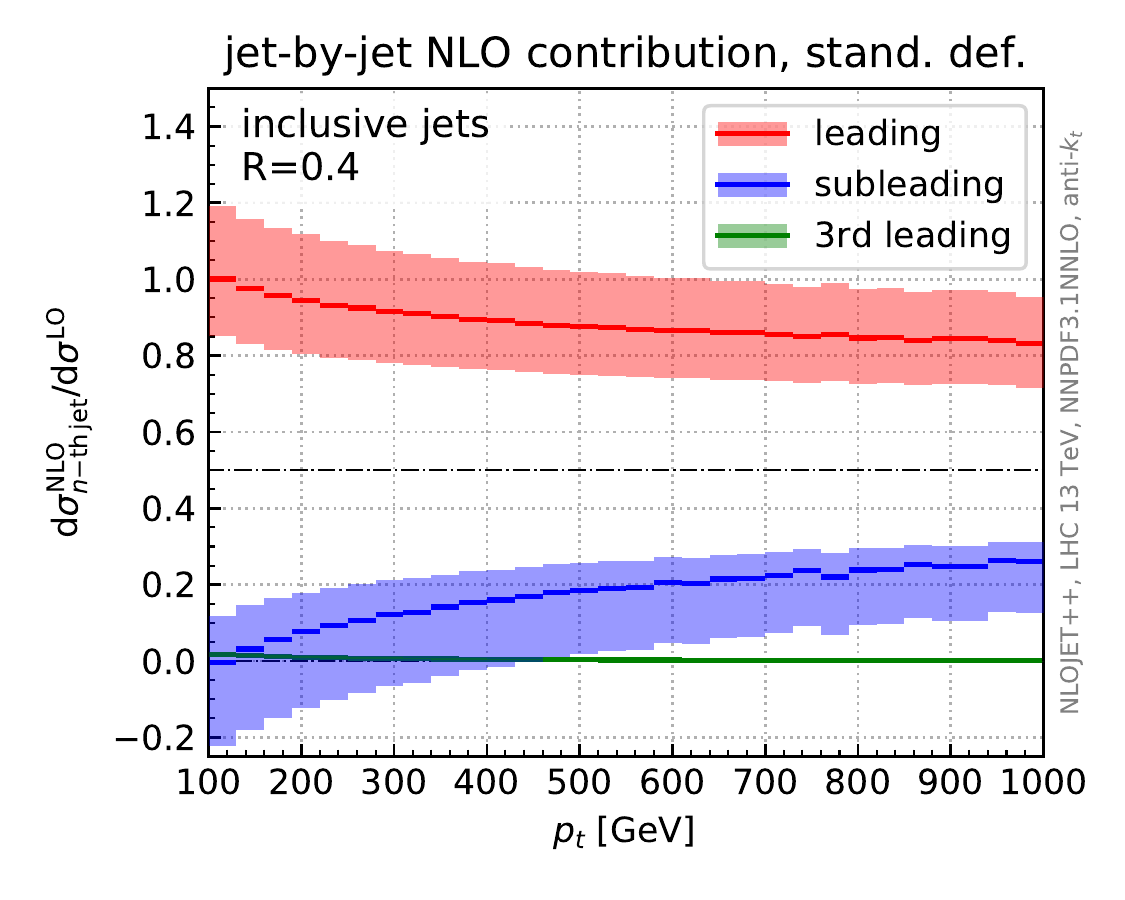}
    \includegraphics[width=0.45\textwidth]{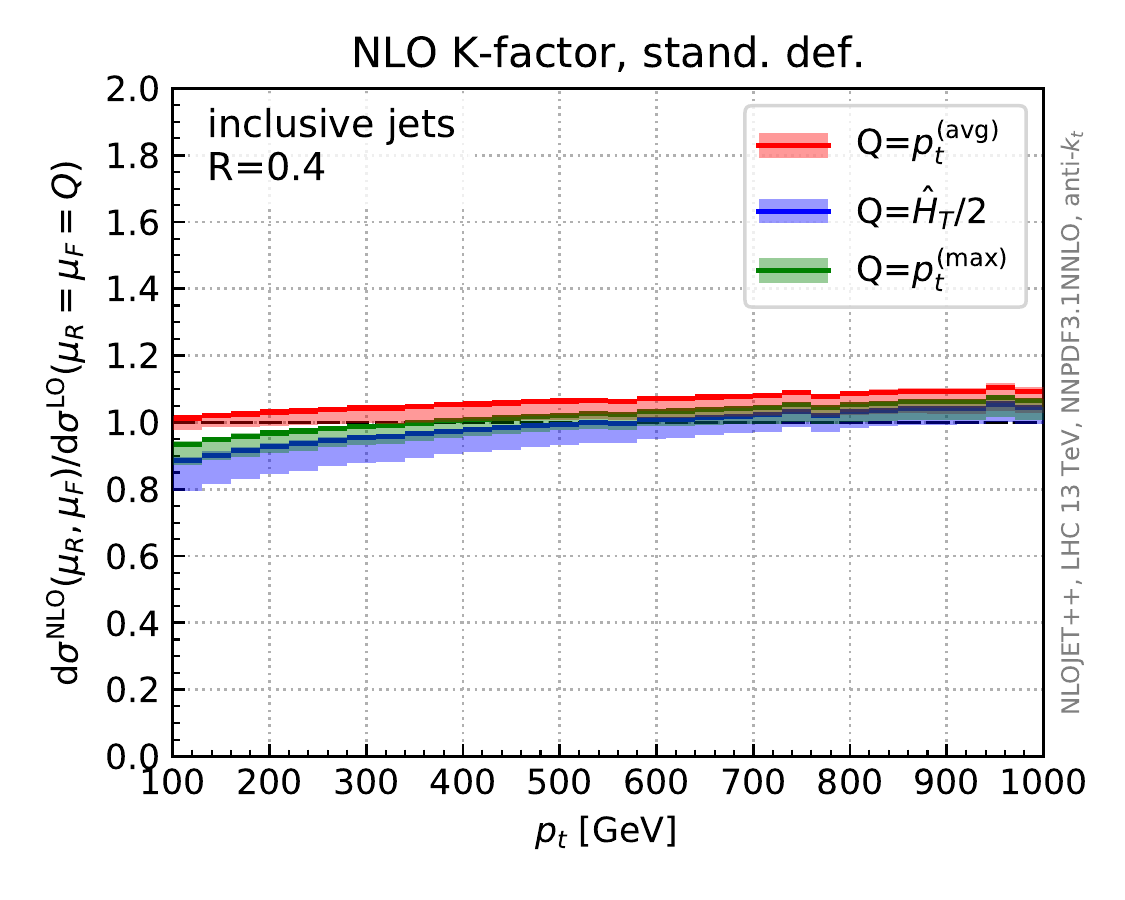}
    \caption{Left: Contributions from the leading, subleading,
      and third-leading jet to the NLO inclusive cross-section, with
      central scale choice $\mu_R = \mu_F =\ptt^\text{(avg)}$ Eq.~(\ref{eq:avscale}).
      Right: Inclusive NLO $K$ factors, with three different central
      scale choices (see text).  }
  \label{fig:ptfr}
\end{figure}

\begin{figure}[t]
  \centering
  \includegraphics[width=0.45\textwidth]{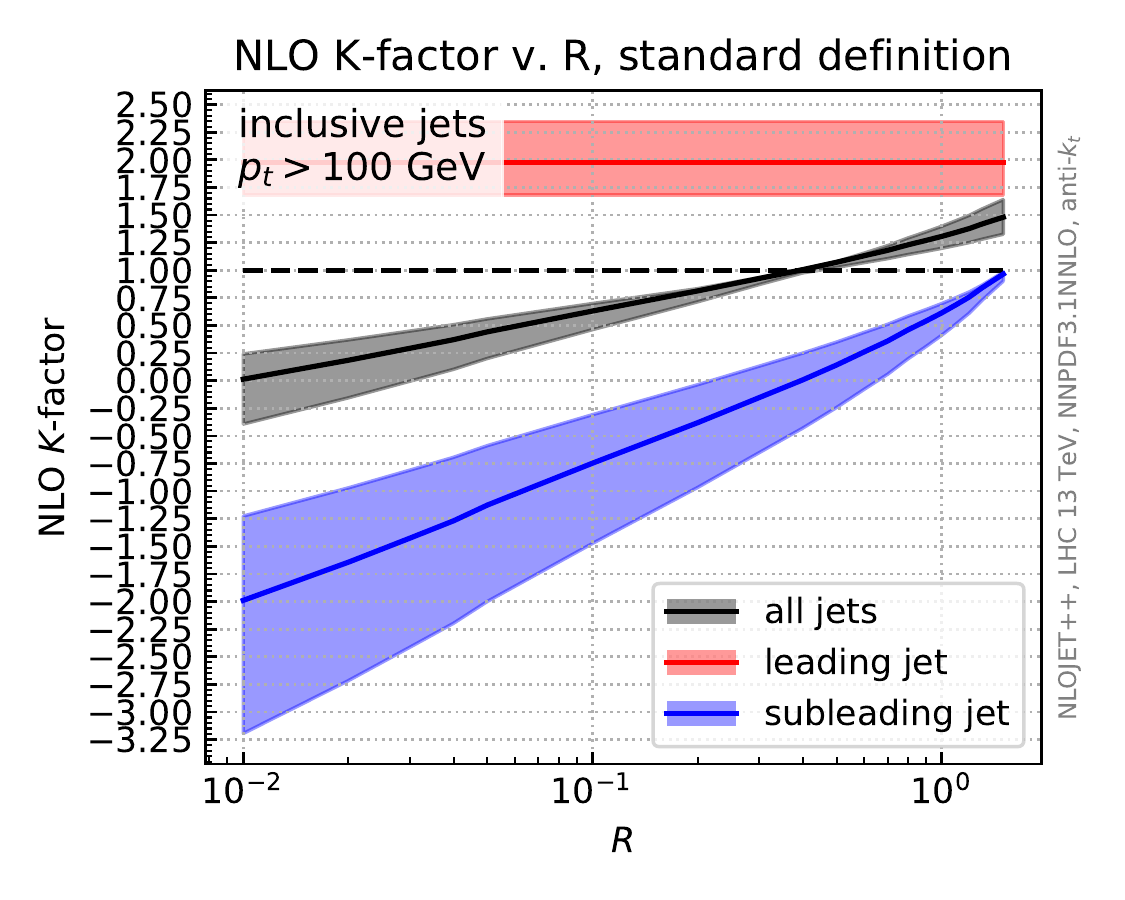}
  \includegraphics[width=0.45\textwidth]{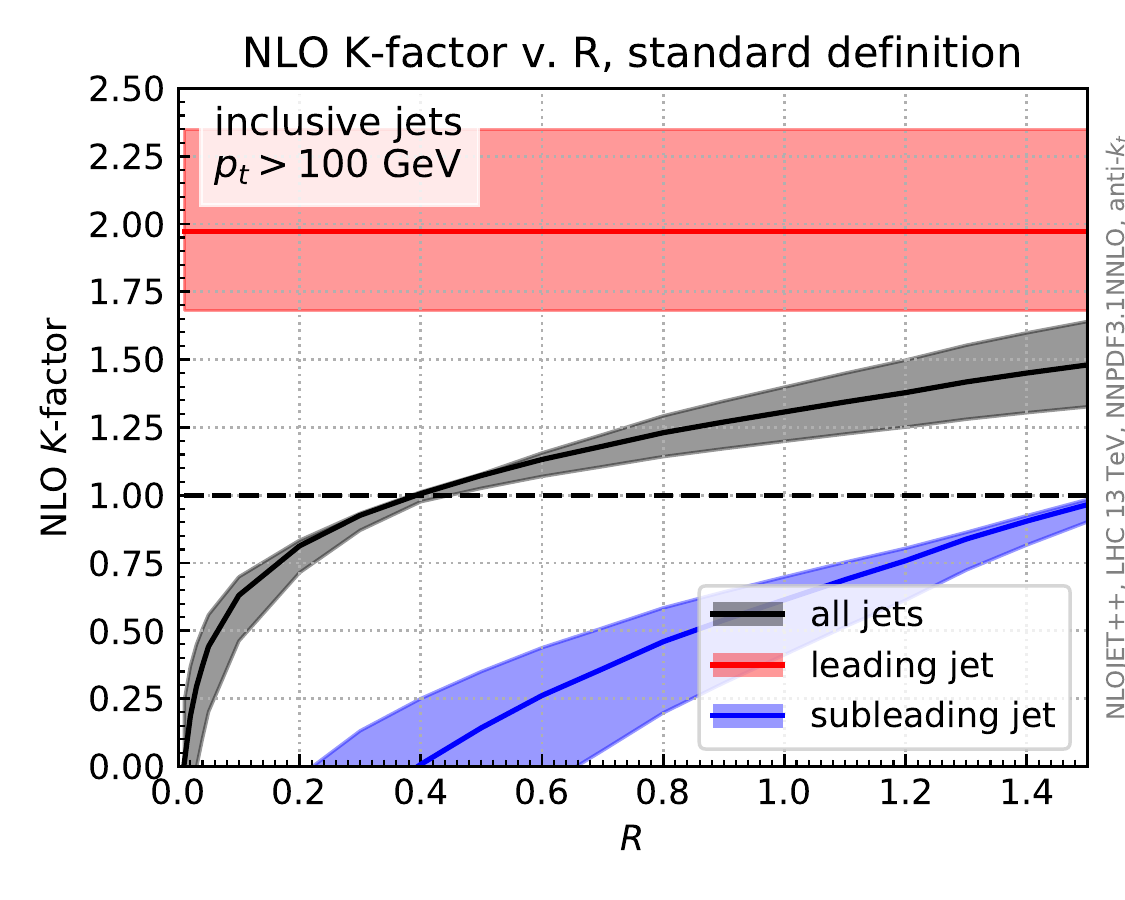}
  \caption{The NLO $K$ factor for the single-inclusive total jet cross-section as
    a function of the radius $R$ of the jet (black). The  contributions from
    the leading jet (red) and the subleading  jet (blue) are also
    shown. Results are plotted both with a logarithmic (left) and
    linear (right) scale.}
  \label{fig:KvR}
\end{figure}
They are shown   in Fig.~\ref{fig:ptfr} for the standard definition.
Three main features are apparent.
First, while the total NLO $K$ factor is
quite close to one (see the right plot in Fig.~\ref{fig:ptfr}), the
individual $K_{n}$ for the leading and subleading jet deviate from
their leading order value, $1/2$, by sizable amounts
(see the left plot in Fig.~\ref{fig:ptfr}). However, they almost
exactly compensate when added up into the total cross-section,
yielding a total NLO $K$ factor close to 1, as well as a scale
uncertainty much smaller than those of the individual $K_n$.
This almost exact compensation is largely accidental as it depends on
the value of the jet radius. This can be seen in Fig.~\ref{fig:KvR},
where we plot the $K$ factor for the total cross-section as a function
of $R$: the leading and the second leading jet $K$ factors only
compensate (up to a residual $\sim 10\%$ effect) in the region
$R \sim 0.3-0.6$.  This effect has also been noticed in
Refs.~\cite{Dasgupta:2016bnd,Bellm:2019yyh}.

The behavior of the
individual jet $K$ factors can be explained in a simple fashion.  At
NLO, the $K$ factor of the leading jet $K_1$ is substantially larger
than one, most likely a consequence of recoil effects amplified by the
fact that the LO cross-section is steeply falling --- typically with a
power around 5 --- in $\ptt$. Furthermore, at NLO, $K_1$ does not
depend on $R$, as explicitly visible in Fig.~\ref{fig:KvR} and as we
show analytically in Sect.~\ref{sec:angen} below.  However $K_2$
decreases at small $R$ since out-of-cone final state radiation depends
on the jet radius and has the effect of lowering the $\ptt$ of the
emitter. This effect is again drastically enhanced by
the steeply-falling nature of the LO differential cross-section in
$p_t$.

It can be seen from the logarithmic scale that the dependence of the
cross-section on $\ln R$ becomes linear only for $R\lsim 0.2$: hence,
the logarithmic contribution dominates the cross-section only in the
very small $R$ region, and indeed resummation was shown to be necessary in
this region in
Refs.~\cite{Dasgupta:2016bnd,Kang:2016mcy,Liu:2017pbb}.
For larger $R$ the $\ln R$ term is still
sizable, but the bulk of the $\ln R$ effects is captured by the exact
NLO result, and for $R\gsim 0.4$ there is a modest benefit in
resumming them, as also shown in
Refs.~\cite{Dasgupta:2016bnd,Kang:2016mcy,Liu:2017pbb}, where this
resummation was performed explicitly. 

% In addition, recoil effects from initial
% state radiation lifts the born level $\ptt$ degeneracy. This has
% the effect of increasing the $\ptt$ of a jet that then becomes the
% leading one (hence further lowering the $p_t$ of the subleading jet).
% This is a kinematic effect that does not depend on the jet radius and
% which we believe is the main reason for the large positive $K_1$.

%
Second, while the leading and second jet account for
most of the cross-section, the contribution of the third jet to the
total $K$ factor is much smaller (giving a correction of less than 2\% of the LO
cross-section) and almost completely negligible. The dominance of
the first two jets as $\ptt$ grows
is important in determining the qualitative
features of the standard definition, in comparison to the various
other definitions that we consider below. It persists at NNLO,
as shown in Ref.~\cite{Currie:2018xkj}, and it is in fact to be
expected to persist to all orders, as a consequence of the dominance of soft
radiation which, combined with the transverse-momentum conservation,
favours configurations in which two hard jets are back-to-back while all the
others are softer.

Finally, by inspecting the uncertainty bands shown in Fig.~\ref{fig:ptfr}, one
can see that scale variation bands for $R=0.4$ for
different central scale choices do not overlap in the small $\ptt$
region. An in-depth discussion of this problem and how this
  changes when including even higher order QCD corrections is
given in Ref.~\cite{Currie:2018xkj}. It is however clear that this  is
a consequence of the accidental compensation
of the two leading jets discussed above, which then propagates onto the
scale variation. It follows that theoretical uncertainties obtained by
performing standard scale variation for fixed $R\sim0.4$ are
unrealistically small. A more reliable estimate can be obtained
performing  uncorrelated scale
variation~\cite{Dasgupta:2016bnd,Bellm:2019yyh}, which then leads to overlapping scale uncertainties across the
whole $\ptt$ spectrum, analogously to what happens in the context of
jet vetoing, where decorrelated scale variation also leads to  more
realistic uncertainty estimates in the presence of
cancellations~\cite{Stewart:2011cf}.

All this shows that the putative perturbative instability of the
standard definition is in fact a byproduct of an entirely accidental
cancellation which happens only at NLO in a given $R$ range. Because
this cancellation is not protected by a symmetry, one should not
expect it to persist
with a different definition or at higher perturbative orders.

\subsection{Weighted (unitary) definitions}
\label{sec:numerics-wgt}
We now turn to the study of the weighted (unitary) definitions of the
single inclusive-jet cross-section introduced in
Sect.~\ref{sec:definition}.
We start our discussion with case (A), in which a $\ptcut$ is adopted,
and we show that in fact this unitary definition appears to display a
somewhat problematic behavior, whose origin is discussed analytically
in Sect.~\ref{sec:analytics}. We then turn to cases (B) and (C) which
provide a natural way to alleviate this problematic behavior.

\begin{figure}[t]
  \centering
  \includegraphics[width=0.45\textwidth]{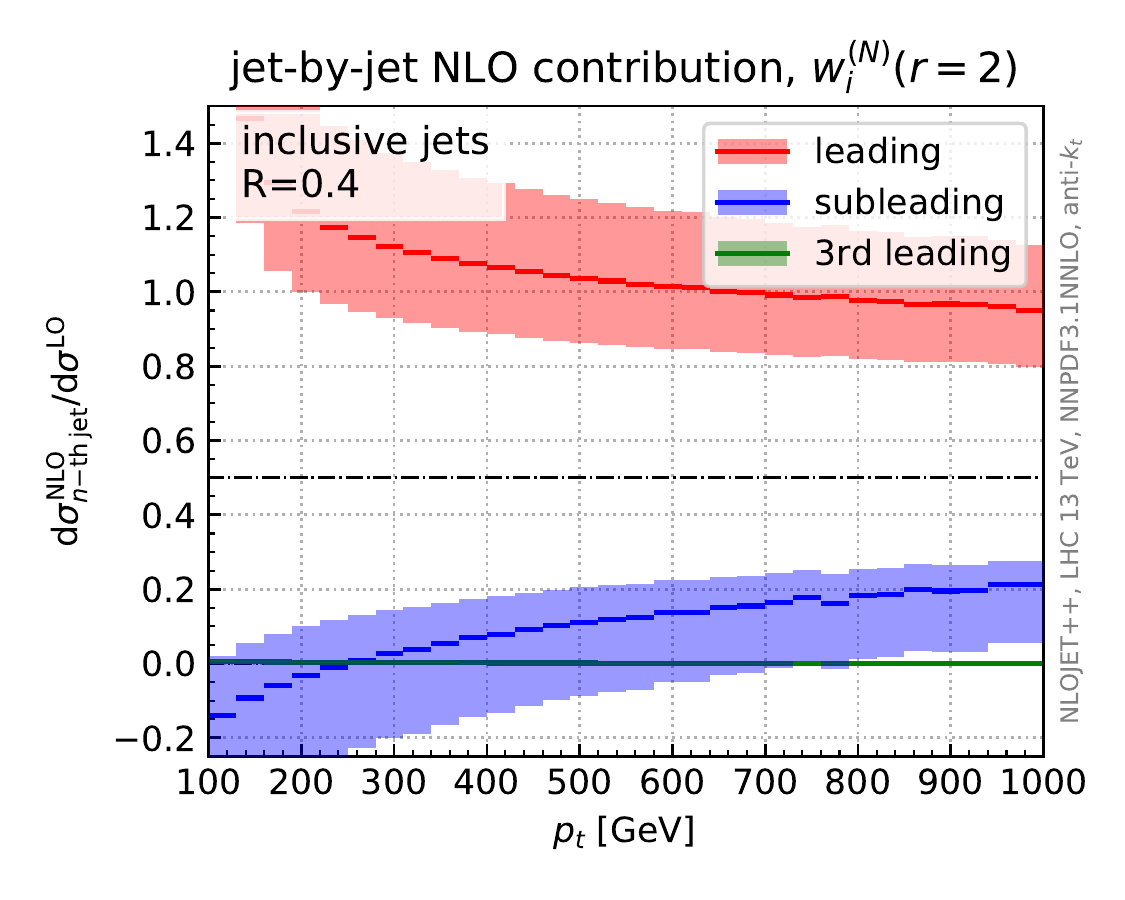}
  \includegraphics[width=0.45\textwidth]{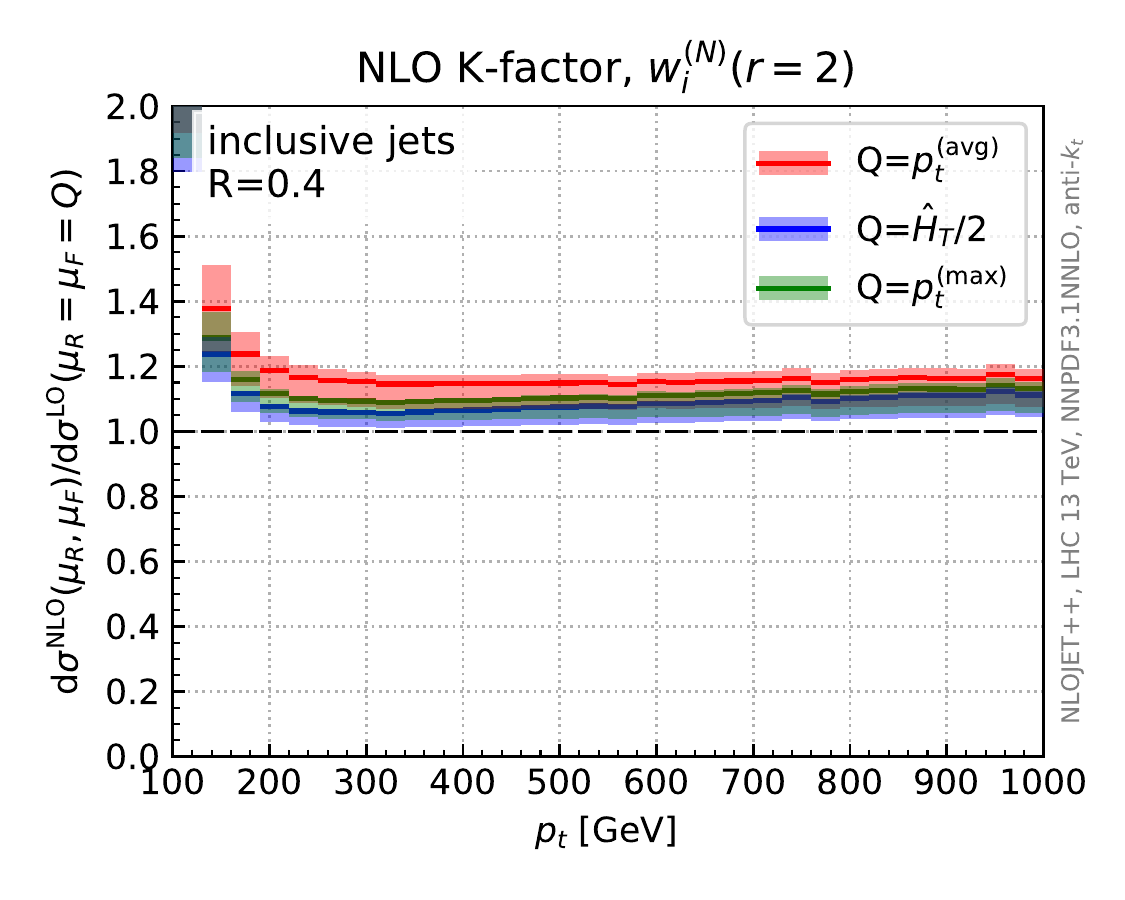}\\
  \includegraphics[width=0.45\textwidth]{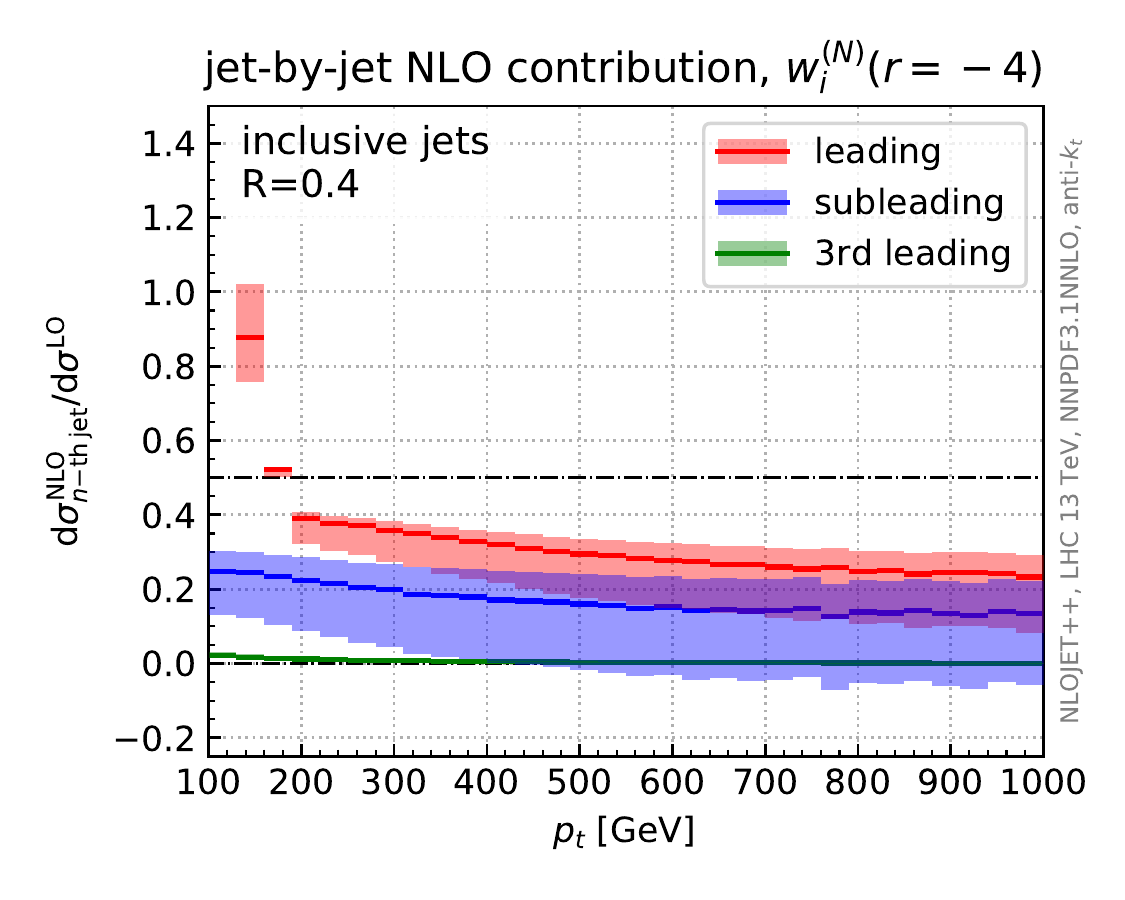}
  \includegraphics[width=0.45\textwidth]{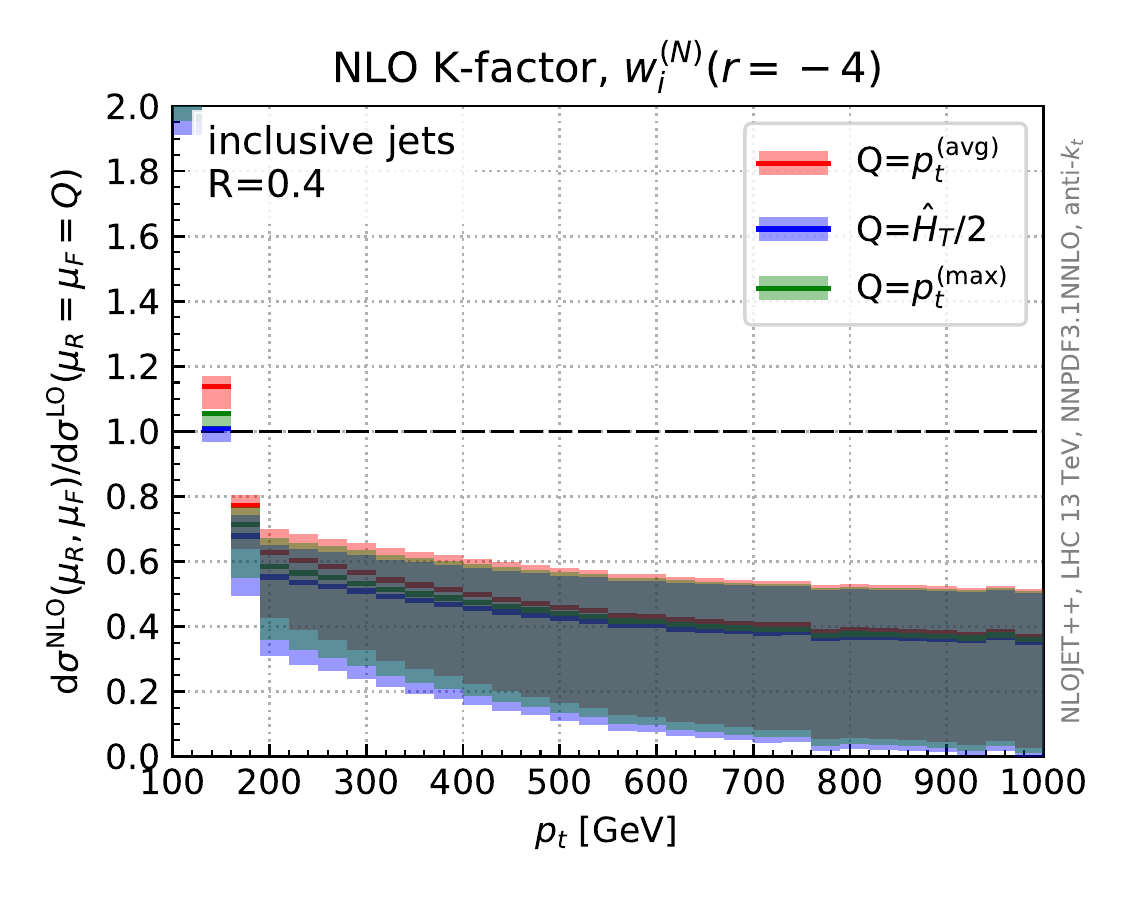}
  \caption{Same as Fig.~\ref{fig:ptfr} but using the weighted
    definitions of type A (see text:
    jets above $\ptcut$) for $r=2$ (top) and $r=-4$ (bottom). }
  \label{fig:ptfr-weighted}
\end{figure}

\paragraph{A. Jets above $\ptcut$.}

In Fig.~\ref{fig:ptfr-weighted} we show again the individual jet
contributions and $K$ factor, now using weighted definitions of type
(A), with  a positive ($r=2$) and a negative ($r=-4$) value for the
exponent in the weights.
Note that the $K_n$, and hence the total $K$ factor, are normalized to the LO
weighted jet cross-section which is exactly half of the LO jet
cross-section obtained with the standard definition. Indeed, at LO we
have $w_1=w_2=1/2$, by kinematic constraint, for the weighted
definition, independently of $r$.
%
% We see that the ($R$-dependent) compensation between the leading and
% second jet $K_n$ that was seen with the standard definition has now
% disappeared. 

We first discuss the behaviour for $\ptt$ far above $\ptcut$.
Broadly speaking, positive weights enhance the difference
between leading and second leading jets, with features that resemble
those of the standard definition for the individual $K_n$ factors.
This is also true, in particular, for the total $K$ factor for $\ptt$
sufficiently larger than $\ptcut$ (top row
of Fig.~\ref{fig:ptfr-weighted}).
Negative values of $r$, on the other hand, have
the effect of balancing the difference between leading and subleading
jets. This results in more similar individual $K_n$ factors, at the
price of an overall larger total $K$ factor (bottom row
of Fig.~\ref{fig:ptfr-weighted}).
At very large $\ptt$ this effect becomes very large, which can be easily understood as follows:
whenever we have three jets passing the $p_t$ cut with
$\pt{1,2}\gg \pt{3}$
% --- which is increasingly likely at higher $p_t$---
we have
\begin{align}
  w_{1,2}^{(3)}(r<0) & =
  \frac{\pt{1,2}^r}{\pt{1}^r+\pt{2}^r+\pt{3}^r}\sim  \left(\frac{\pt{3}}{\pt{1,2}}\right)^{|r|}\ll 1\,,\label{eq:wgt-def-ptcut-largew12}\\
  w_{3}^{(3)}(r<0) & =
  \frac{\pt{3}^r}{\pt{1}^r+\pt{2}^r+\pt{3}^r}\sim 1\,.
\end{align}
The contributions of the two leading jets to the inclusive
 cross-section, which are strongly dominating the NLO cross-section
for the standard definition (or for the weighted definition with
$r\ge 0$), are now power suppressed by the weights.
Furthermore, corresponding virtual corrections
have two jets in the final state with
$w_{1,2}^{(2)}(r<0)= 1/2$. At large $p_t$ real and virtual
corrections with $\ptcut\ll\pt{3}\ll \pt{1,2}\sim p_t$ therefore yield, after
integration over $\pt{3}$, a
negative contribution enhanced by $\log(p_t/\ptcut)$, corresponding to
the large corrections seen in Fig.~\ref{fig:ptfr-weighted}.

Now turning to the region where $p_t\to \ptcut$, we see from
Fig.~\ref{fig:ptfr-weighted} that this weighted definition
(for both positive and negative $r$) develops a singular behavior.
The origin of this behavior is explained analytically in
Section~\ref{sec:analytics}.
For the time being, we note that these
singularities, both for $p_t\gg\ptcut$ and for $p_t\to\ptcut$, are of logarithmic origin and could in principle be
dealt with resummation.

In summary, the weighted definitions of type (A) (with $\ptcut$)
have the undesirable feature of
developing problematically unstable behaviors for $p_t$ close to the
$p_t$ cut as well as at large $p_t$ for $r<0$. In the other $p_t$
regions their perturbative behavior now shows large $K$ factors also
at NLO since the accidental cancellation of the standard definition is spoiled;
while this is perhaps more natural, it does not suggest an improvement
in perturbative behavior over the standard definition.

%\subsection{Weighted (unitary) definitions}
%\label{sec:weighted-all-jets}

\begin{figure}[t]
  \centering
  \includegraphics[width=0.45\textwidth]{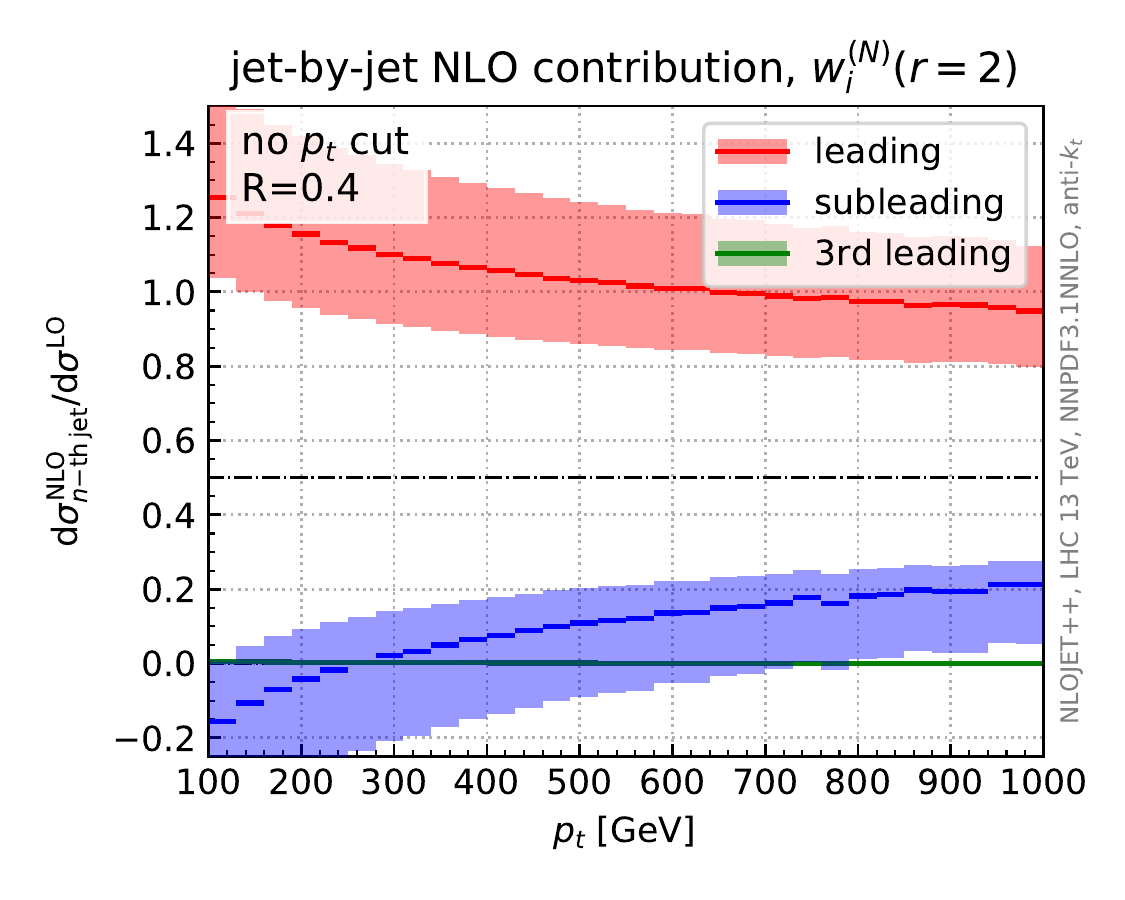}
  \includegraphics[width=0.45\textwidth]{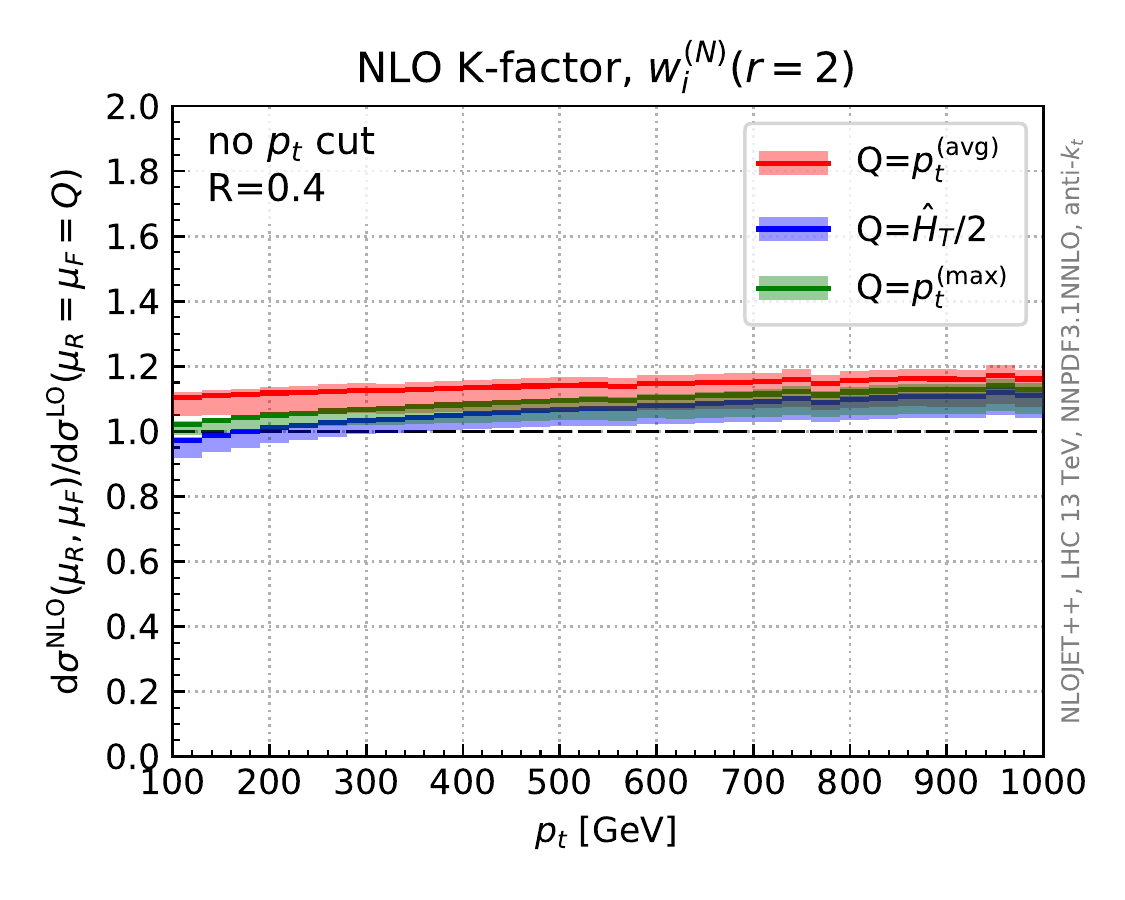}\\
  \caption{Same as Fig.~\ref{fig:ptfr}, but using  a weighted definition of
    type B  (all jets) for $r=2$. }
  \label{fig:ptfr-weighted-all}
\end{figure}

\paragraph{B. All jets.}

A natural way of curing the logarithmic divergence observed when
$p_t\to \ptcut$ using weights of type  (A) is to include all jets down to a $\ptt$ much
smaller than the first bin of the distribution.
Based on Fig.~\ref{fig:ptfr-weighted}, taking a $\ptcut$ two or three
times smaller than the first bin of the distribution would already
get rid of most of the sensitivity to $\ptcut$, e.g.\ without any need for an
additional resummation.
One can view the weighted definition of type (B) as simply taking the
limit $\ptcut\to 0$ and one should not expect our conclusions to
change as long as $\ptcut$ remains much smaller than the first bin of
the distribution, say $\ptcut\sim 20-30$~GeV. 
% 
% In practice, this is
% done by setting $\ptcut=0$, which correspond to the weighted
% definitions of type (B): indeed, our conclusions do not
% change substantially even for $\ptcut\sim 20$~GeV.
%
This possibility is
only sensible for positive weights, for which the
low $\ptt$ part of the spectrum is suppressed. For negative weights  this choice is infrared unsafe.
 
Results are shown in Fig.~\ref{fig:ptfr-weighted-all} for $r=2$.  As
expected, the singular behavior of the $K$ factor for $p_t$ close to
$\ptcut$ is now absent, and features similar to those of the standard
definition are now recovered. Specifically, non-overlapping scale variation
bands are observed in the low $\ptt$ region, though to a smaller
extent than in the standard case.  As a last comment, we
have checked that this definition does not suffer from large
non-perturbative corrections, such as those coming from underlying
events, despite involving low-$\ptt$ jets. In a practical
  experimental context, one would still need to make sure that this
  remains true with realistic pileup conditions.

\paragraph{C. Two leading jets.}

\begin{figure}[t]
  \centering
  \includegraphics[width=0.45\textwidth]{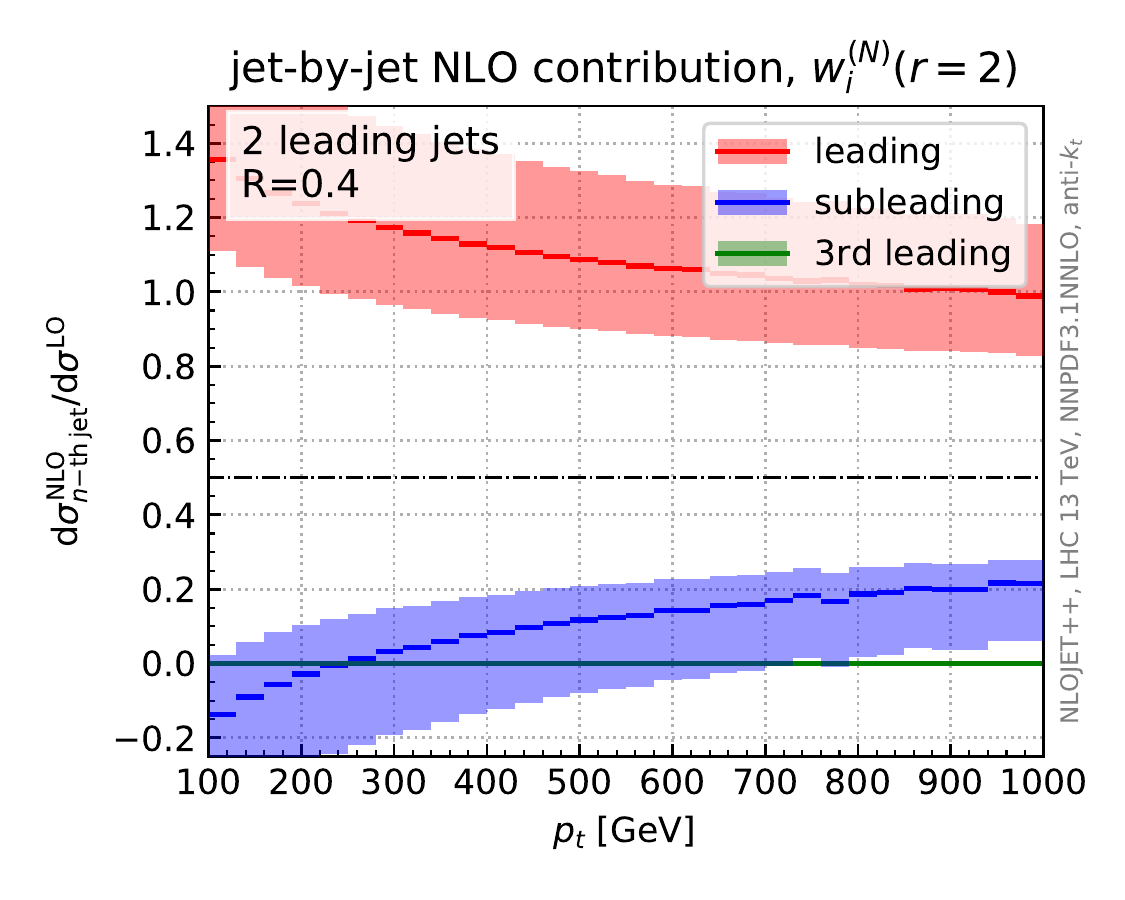}
  \includegraphics[width=0.45\textwidth]{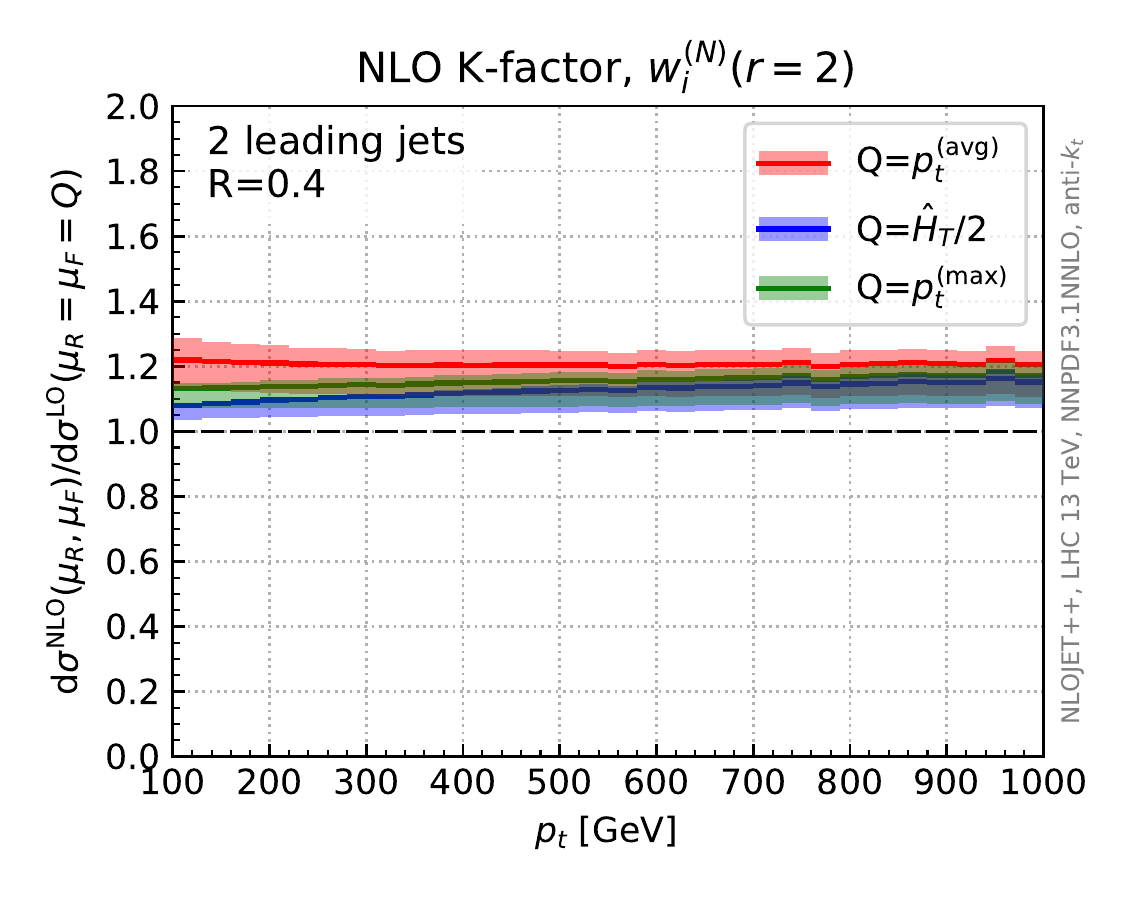}\\
  \includegraphics[width=0.45\textwidth]{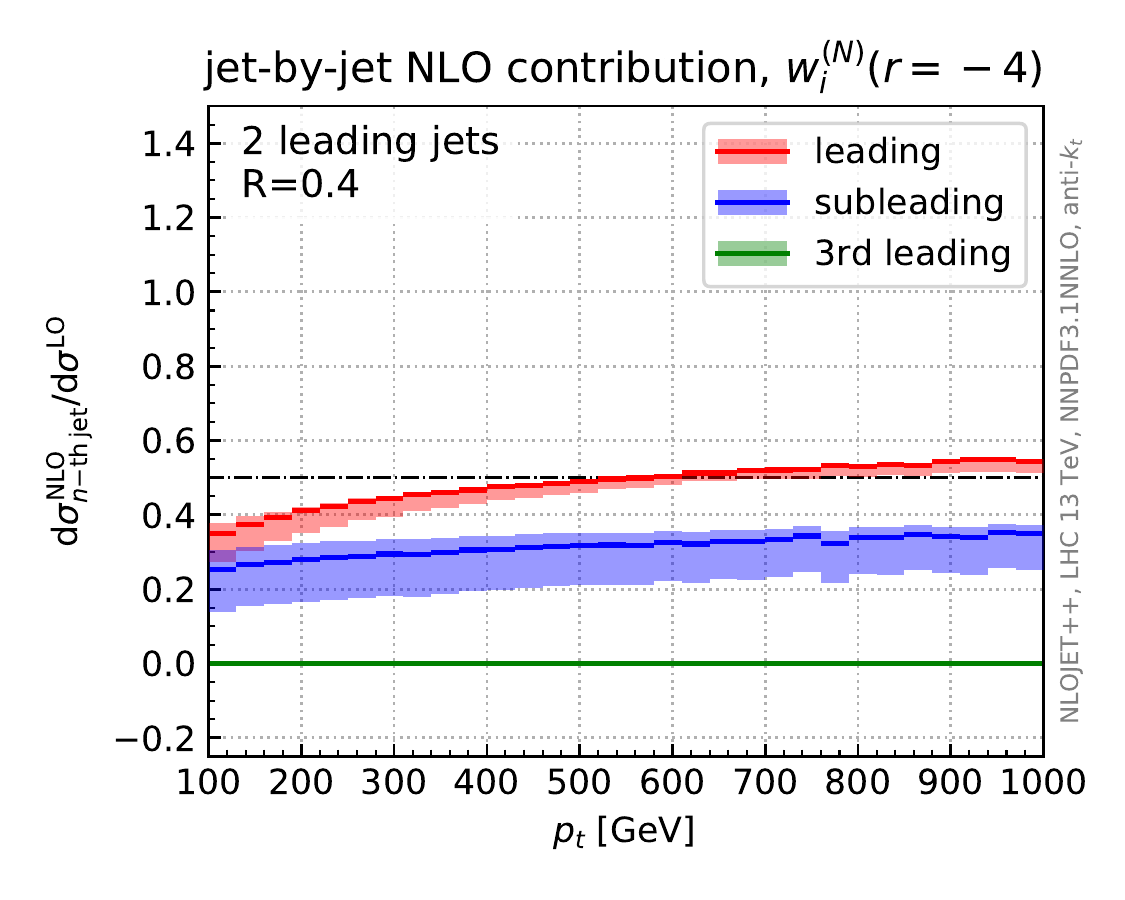}
  \includegraphics[width=0.45\textwidth]{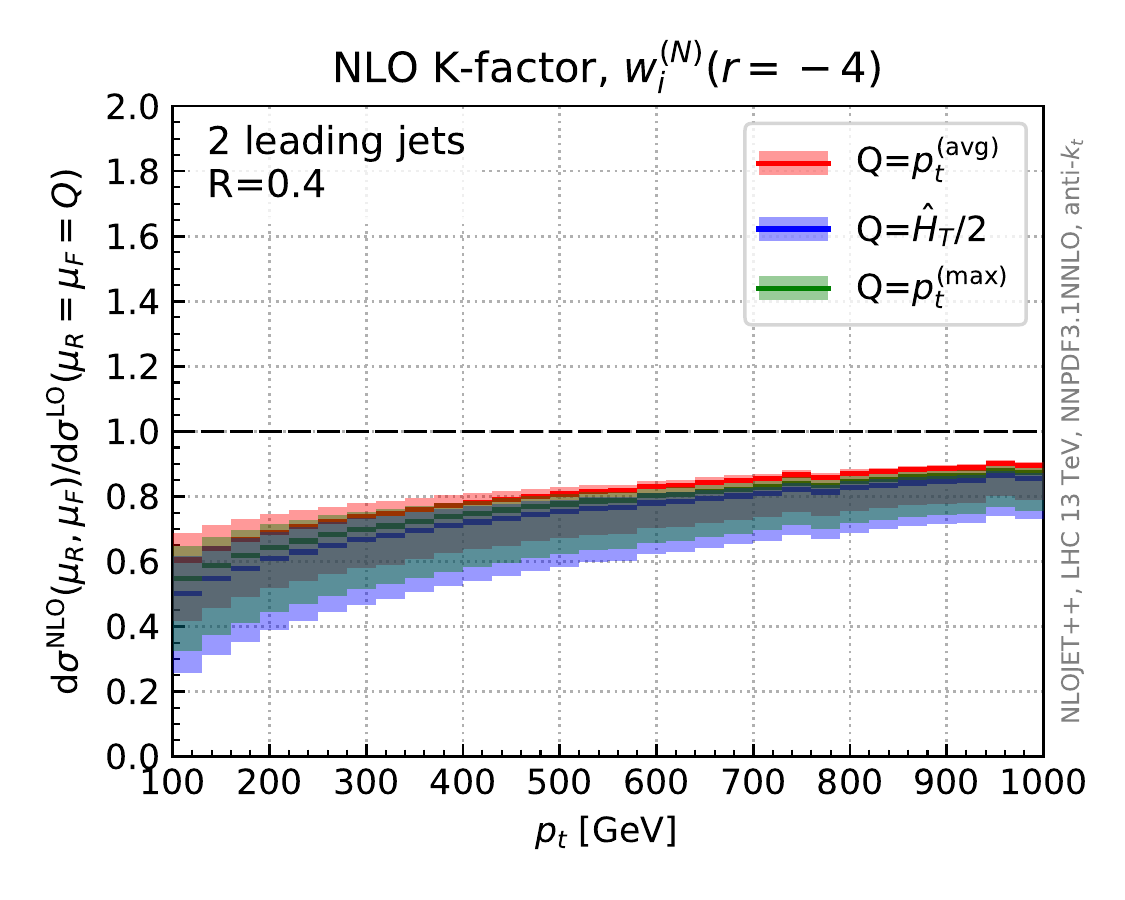}
  \caption{Same as Fig.~\ref{fig:ptfr}, but using the weighted
    definitions of type C (two leading jets) for $r=2$ (top) and $r=-4$ (bottom). 
} \label{fig:ptfr-weighted-2leading}
\end{figure}

An alternative choice, motivated by the observation that the contribution of the
third jet to the inclusive jet cross-section is much smaller than that
of the first two jets
(see Fig.~\ref{fig:ptfr}) is to switch to definitions of type (C), in
which only the two leading jets are included in the weights, whether
or  not they pass a given $\ptcut$. Clearly this should also remove
the problem of the behavior  for $\ptt\sim\ptcut$ of definitions of
type (A).
This approach is similar in
spirit to what is done when looking at the dijet cross-section.
Results in this case  are presented in
Fig.~\ref{fig:ptfr-weighted-2leading} for the individual $K$ factors
$K_n$ and the total $K$ factor.
The situation for positive $r$ is
again similar to what we observe for the standard definition: in
particular there seems to be a large compensation between the
leading and subleading jets, leading to a rather flat $K$ factor,
though  larger
than in the standard case.

As explained above, negative values of $r$
have the effect of normalizing the
individual $K_n$ factors for the leading and subleading jets, reducing
the effect of the compensation seen in the standard
case.
Furthermore, the uncertainty bands obtained for the three different
scale choices now overlap.
Nevertheless, the inclusive $K$ factor is relatively larger than
for the standard definition and shows a
somewhat strong $\ptt$ dependence.

Comparing these results to the other weighted definitions, we see that
the logarithmic divergence for $p_t$ close to $\ptcut$ which is
observed in Fig.~\ref{fig:ptfr-weighted} when using the jets above
$\ptcut$ has now disappeared for both positive and negative $r$.
This, as discussed above, is expected: the weights do not depend on 
whether one or two of the two leading jets passes the
$\ptcut$, so the definition becomes independent of the cut.
Furthermore, the issue with large $K$ factors at large $p_t$
for negative $r$ when including jets above $\ptcut$ has also
disappeared. This is simply because the third jet no longer
contributes to the weights and therefore the large contribution seen
in Eq.~(\ref{eq:wgt-def-ptcut-largew12}) is absent.

In summary, weighted definitions of type (C) behave similarly to the standard definition
for positive $r$. The perturbative behavior for negative $r$ changes,
with some desirable features  (the individual
$K$ factors $K_1$ and $K_2$ are similar, and the scale uncertainty
bands for different scale choices overlap), and some undesirable ones
(the overall $K$ factor is larger).

\section{Non-unitarity and perturbative behavior}\label{sec:analytics}

We now show how several features of the results presented in the
previous section can be understood on the basis of simple analytic
arguments. Specifically, we show that the behavior in the vicinity of
$\ptcut$ is strongly tied to the unitarity, or lack thereof, of the
various definitions.

We first provide general (if somewhat formal) arguments, exploiting the fact that 
at NLO the jet functions used for partitioning the phase-space
have a compact and manageable form.
We then perform more explicit calculations using a soft and collinear
approximation which shows that the effects discussed in Section~\ref{sec:numerics-wgt}
have a simple leading logarithmic origin.

\subsection{Dependence on $\ptcut$ and $R$: a general argument}
\label{sec:angen}
In order to understand the behavior of various definitions we need an
explicit expression for the contribution to the $N$-jet cross-section
$ \frac{d \sigma^{(k)}_{\text{N\,jet}}}{d \ptt}$ introduced in
Eq.~(\ref{eq:sigmaknNjet}) and to the $n$-th jet cross-section
$\frac{d \sigma^{(k)}_{\text{n-th\,jet}}}{d \ptt}$ introduced in
Eq.~(\ref{eq:sigmaknN}).  These can be constructed in terms of
parton-level cross-sections by introducing explicit jet functions that
cluster final-state partons into jets, in the latter case further
supplemented by a function that selects the $n$-th leading jet, and
bins the result into a fixed $\ptt$ bin. In order to cancel infrared
singularities, the $k$-th order contribution must be constructed by
adding up contributions coming from final states with a number of
final-state partons that goes from two (with $k$ virtual
loops), up to $k+2$ (with $k$ real emissions on top of the Born level). For instance the NLO $k=1$ term receives
contributions both from a two-parton final state with one loop, and
from a real emission three-parton state, and so on.

Explicitly, we can write the $N$-exclusive jets contribution,
Eq.~(\ref{eq:sigmaknNjet}), as a sum of terms where the $N$ jets are produced
from an $m$ parton final-state, ${\rm d}\Phi_m$,
\begin{equation}
  \label{eq:njetfunct}
  \frac{{\rm d} \sigma^{(k)}_{N\,\text{jets}}}{{\rm d} \ptt}=  \sum_{m=2}^{k+2}  \int {\rm d}{\Phi}_{m}         
     \frac{{\rm d}\hat{\sigma}^{(k)}_m}{{\rm d} {\Phi}_{m} } G_{m \to {N\,\text{jets}} }({\Phi}_m, \ptt)\,,
\end{equation}
where  $G_{m \to {N\,\text{jets}}}$ is the jet function which
cluster $m$ partons into $N$ jets. $G_{m \to {N\,\text{jets}}}$ contains the function $F_N$, Eq.~(\ref{eq:stw}),
which in turn includes the possible weights. The jet function thus depends on
the jet momentum $\ptt$, and on the partonic phase space variables
${\rm d}\Phi_m$.

We can give an explicit expression of $G_{m\to N}$ at NLO ($k=1$). For
this, let us denote by $k_{ti}$ the parton transverse momenta, with
$\kt{1}\ge \kt{2}\ge \kt{3}$. Using the
anti-$k_t$~\cite{Cacciari:2008gp} jet clustering with
$R<\tfrac{\pi}{2}$, one has
\begingroup \allowdisplaybreaks
\begin{align}
  \label{eq:jetf-v}
  G_{2 \to 1} & = G_{2 \to 3} = 0\\
  \label{eq:jetf-v2}
  G_{2 \to 2} & = \Theta(\ptt>\ptcut)
                \left\{ 2\,w^{(2)}(\ptt;\ptt, \ptt)\, \delta(\ptt - \kt{1}) \right\}\\
  \label{eq:jetf1}
  G_{3 \to 1} & = \Theta(\Delta R_{23}>R)\,
                  \Theta(\kt{1} > \ptcut > \kt{2} > \kt{3})
                \left\{
                w^{(1)}(\ptt;\ptt) \,\delta(\ptt - \kt1)
                \right\}\\
  G_{3 \to 2} & = \Theta(\Delta R_{23}>R)\,
                  \Theta(\kt{1} > \kt{2} > \ptcut > \kt{3})
                \left\{
                \sum_{i=1}^2 w^{(2)}(\ptt;\kt{1},\kt{2})\,\delta(\ptt - \kt{i})
                \right\}
                \nonumber \\
                & + \Theta(\Delta R_{23} < R)\,
                  \Theta(\ptt > \ptcut)\,
                \left\{
                2\,w^{(2)}(\ptt;\ptt,\ptt)\, \delta(\ptt - \kt{1})
                \right\}                
  \label{eq:jetf2}\\
  \label{eq:jetf3}
  G_{3 \to 3} & = \Theta(\Delta R_{23} > R)\,
                \Theta(\kt{1} > \kt{2} > \kt{3} > \ptcut)
                \left\{
                \sum_{i=1}^3 w^{(3)}(\ptt;\kt{1},\kt{2},\kt{3})
                \delta(\ptt-\kt{i})
                \right\},
\end{align}
\endgroup
where we have defined, as is customary, 
%\begin{equation}
$\Delta R_{ij} = \sqrt{(\Delta \phi_{ij})^2 + (\Delta y_{ij})^2}$,
%\end{equation}
as the distance between parton $i$ and parton $j$ in the
rapidity-azimuth plane, with $y$ and $\phi$ the rapidity and the
azimuthal angle respectively.
Note also that, due to momentum conservation, it is sufficient to
consider the recombination of the two softest partons. The second line
of Eq.~(\ref{eq:jetf2}) corresponds to the case where the two softest
partons cluster, yielding two back-to-back jets of momentum $\kt{1}$.

Using Eqs.~(\ref{eq:jetf-v})-(\ref{eq:jetf3}), the issue of unitarity vs. cancellation of
the dependence on $\ptt$ is easily understood.
On the one hand, it is clear that the standard definition is not
unitary and only the weighted definitions are unitary because
\begin{equation}\label{eq:intptwgt}
  \int {\rm d}\, \ptt\,G_{3 \to 1}+  G_{3 \to 2} + G_{3 \to 3} \big|_{\text{wgt}} = \Theta(\kt{1} > \ptcut)\,,
\end{equation}
This result, valid for any $r$,
means that integrating the single-jet cross-section over $\ptt$ yields the total
cross-section for producing (at least) one jet above $\ptcut$ (with
definitions of type (A) in the sense of Sect.~\ref{sec:definition}) or the total cross-section (for definitions of
type (B) or of type (C)). Hence these choices are unitary, and thus
the standard choice cannot be.

On the other hand, it is clear that the inclusive cross-section is
independent of $\ptcut$ when using the standard definition. Indeed, in
this case one has
%only when all weights are set to one. Indeed,
%using the standard definition we see that
\begin{equation}\label{eq:totnu}
  G_{3 \to 1}+  G_{3 \to 2} + G_{3 \to 3} \big|_{\text{std}}
  = \Theta(\ptt > \ptcut)
    \Bigg\{
      \Theta(\Delta R_{23} > R)
    \left[
      \sum_{i}^3\delta(\ptt - \kt{i})
    \right] 
  +
      \Theta(\Delta R_{23} < R)
    2\,\delta(\ptt - \kt{1})
  \Bigg\},
\end{equation}
where now the subscript ``std'' denotes that in the definition of
$F_N$, Eq.~(\ref{eq:stw}), the standard case in Eq.~(\ref{eq:wexpr})
has been selected.  The result Eq.~(\ref{eq:totnu}) is manifestly
independent of $\ptcut$ since all the dependence on $\ptcut$ is
factored in an overall $\Theta$ function which is always satisfied as
long as one has at least one jet in the event.
In practice, the dependence on $\ptcut$ disappears since, when
integrating over the partonic transverse momenta, the $N$-jet
contribution has $\ptcut$ as a lower bound of integration while the
$N-1$-jet contribution has $\ptcut$ as an upper bound. When summing
both contributions, the $\ptcut$ dependence cancels.
%
% What happens here is that while  each of the $N$-jets cross-sections
% depends on $\ptcut$, because it is
% obtained by integrating up the differential parton-level cross-section
% with respect to the transverse momentum with $\ptcut$ as a lower
% limit, this dependence disappears in the sum since the $N-1$-jet
% contribution has $\ptcut$ as an upper limit of integration, and so on.
%

When one instead uses a unitary definition which explicitly introduces
a $\ptcut$ dependence, such as definition (A), this cancellation is
spoiled: whether a jet passes a cut or not changes the weights of all
the other jets, thereby introducing a cutoff dependence of the
observable. The lack of cancellation then propagates into the
individual $n$-th jet cross-sections, thus explaining the singular
behavior observed in Fig.~\ref{fig:ptfr-weighted} when
$\ptt\sim\ptcut$.
% 
% This cancellation only takes place using a non-unitary definition
% because it is only with a non-unitary definition that the weight
% carried by one jet in the cross-section does not depend on how many
% other jets are there. With a unitary, weighted definition
% depending on $\ptcut$, such as definition (A),
% whether a jet passes a cut or not changes the weights of all the other
% jets, thereby introducing a cutoff dependence of the observable.  Lack
% of cancellation with the weighted definitions then propagates into the
% individual $n$-th jet cross-sections, thus explaining the singular
% behavior observed in
% Fig.~\ref{fig:ptfr-weighted} when
% $\ptt\sim\ptcut$.
%
Of course this cutoff dependence is not present for the two other
weighted definitions, (B) and (C), even if the weight associated to a jet
still depends on the other jets in the event,
which is needed to eventually ensure the unitarity of the cross-section.

We can similarly understand the $R$ dependence or lack thereof of the
leading jet contribution, which as discussed in
Sect.~\ref{sec:numerics-std} controls the behavior of
the NLO $K$ factor, by introducing explicit expressions for individual
jet functions. We now need to consider
the $n$-th leading jet contribution, Eq.~(\ref{eq:sigmaknN})
\begin{equation}
\label{eq:nthjetfunct}
  \frac{{\rm d} \sigma^{(k)}_{n\text{-th\,jet}}}{{\rm d} \ptt}=\sum_{m=2}^{k+2}  \int {\rm d}{\Phi}_{m}         
  \frac{{\rm d}\hat{\sigma}^{(k)}_m}{{\rm d} {\Phi}_{m} }  S_{m \to {n\text{-th jet}}}({\Phi}_m, \ptt)\,,
\end{equation}
where the functions $S_{m \to {n\text{-th jet}}}$ are defined summing
the contributions coming from the $n$-th jet in the functions $G$
given above. By direct calculation, we find
\begingroup
\allowdisplaybreaks
\begin{align}
  \label{eq:S2}
  S_{2 \to \pt{1}}
  & =  S_{2 \to \pt{2}}
    = \frac{1}{2}\,G_{2\to 2}\\
    S_{3 \to \pt{1}}
    &=
     \Theta(\ptt > \ptcut)\, \delta(\ptt-\kt{1})
    \,\biggl\{
       \Theta(\Delta R_{23} > R)\, \Bigl[
      \Theta(\ptcut > \kt{2} > \kt{3})w^{(1)}(\ptt;\kt{1}) \nonumber \\
  & \hphantom{=\Theta(\ptt > \ptcut)\, \delta(\ptt-\kt{1})    \,\biggl\{ \Theta(\Delta R_{23} > R)\, \Bigl[}
       +\,\Theta(\kt{2} > \ptcut > \kt{3})\, w^{(2)}(\ptt;\kt{1},\kt{2})\nonumber\\
  &\hphantom{=\Theta(\ptt > \ptcut)\, \delta(\ptt-\kt{1})    \,\biggl\{ \Theta(\Delta R_{23} > R)\, \Bigl[}
    + \Theta(\kt{2} > \kt{3} > \ptcut) w^{(3)}(\ptt;\kt{1},\kt{2},\kt{3})\Bigr]\,  \nonumber\\
    &\hphantom{=\Theta(\ptt > \ptcut)\, \delta(\ptt-\kt{1})    \,\biggl\{}+ 
    \Theta(\Delta R_{23} < R) \, w^{(2)}(\ptt;\ptt,\ptt) \biggr\}  \label{eq:S3pt1}
    \\
  \label{eq:S3pt2}
    S_{3 \to \pt{2}}
    &= \Theta(\kt{1} > \ptt > \ptcut)
       \biggl\{\Theta(\Delta R_{23} > R)\,\delta(\ptt - \kt{2})
       \Bigl[\Theta(\ptcut > \kt{3})\, w^{(2)}(\ptt;\kt{1},\kt{2})
      \nonumber \\
     &\hphantom{= \Theta(\kt{1} > \ptt > \ptcut)
       \biggl\{\Theta(\Delta R_{23} > R)\,\delta(\ptt - \kt{2})
       \Bigl[}
       + \Theta(\kt{3} > \ptcut)\, w^{(3)}(\ptt;\kt{1},\kt{2},\kt{3}) \Bigr]\,
    \nonumber\\
    &\hphantom{= \Theta(\kt{1} > \ptt > \ptcut)
       \biggl\{}+ 
    \Theta(\Delta R_{23} < R)\,\delta(\ptt - \kt{1}) \, w^{(2)}(\ptt;\ptt,\ptt) \biggr\}
    \\
  \label{eq:S3pt3}
  S_{3 \to \pt{3}}
    &=
     \Theta(\kt{1} > \kt{2} > \ptt > \ptcut)\, \delta(\ptt-\kt{3})
     \,
       \Theta(\Delta R_{23} > R)\, w^{(3)}(\ptt;\kt{1},\kt{2},\kt{3})\,.
\end{align}
\endgroup
If now one sets all weights
$w=1$, Eq.~\eqref{eq:S3pt1} takes the form
\begin{equation}
  \label{eq:Slj}
S_{3 \to \pt{1}}\big|_{\text{std}} = \Theta(\ptt > \ptcut)\,
\delta(\ptt-\kt{1})=   S_{2 \to \pt{1}}\big|_{\text{std}}\,,
\end{equation}
where the subscript ``std''  again denotes that in the definition of $F_N$,
Eq.~(\ref{eq:stw}), the standard case in Eq.~(\ref{eq:wexpr}) has been selected. 
This means that all the $\Theta$ functions simplify, leading to an
overall factor providing a condition that is always 
satisfied if at least one jet in the event is above $\ptcut$.
At NLO, the leading jet contribution is therefore always given by the
transverse momentum of the hardest parton (this is valid for both the
real contribution with three partons in the final state and the virtual
corrections with two partons), independently of the jet
radius $R$.
Note that, one can similarly see that for any weighted definition, at
NLO, corrections to the leading jet are $R$-dependent for the same
reason that the weighted definitions depend on $\ptcut$: the value of
the weights depend on how many partons have $\Delta R_{ij}>R$.
Furthermore, the NLO corrections for the subleading and third-leading
jet also depend on $R$. This is trivial for the latter which shows an
explicit $R$ dependence in~(\ref{eq:S3pt3}). For the subleading jet,
this is due to the fact that the $p_t$ of the jet changes (between
$\kt{1}$ and $\kt2$) depending on how $\Delta R_{23}$ compares to $R$.

% \gs{I tink we do not need this paragraph anymore, right?}
% %
% This $R$ independence of the NLO corrections to the leading jet can be
% understood physically in the following way. On the one hand, virtual corrections only affect the
% LO kinematics, where two partons are clustered into two jets, which is
% $R$ independent for both the leading or the subleading jet,
% Eq.~\eqref{eq:S2}. On the other hand, real emission contributions to
% the leading jets are $R$ independent because of two reasons.
% First, the momentum of the leading jet always coincides
% with the momentum of the leading parton (if $R$ is small, the three partons
% are resolved individually; if $R$ is large, the second and third leading partons
% are clustered together and this jet has the same momentum as the jet
% formed by the leading parton). This is why the delta function in Eq.~\eqref{eq:S3pt1}
% can factorise. Second, and consequently, the theta functions inside the curly bracket
% in Eq.~\eqref{eq:S3pt1} are a partition of unity, because there is always a leading jet,
% and then the $R$ dependence disappears.
% This does not happen neither for the subleading jet (as the delta function
% in Eq.~(\ref{eq:S3pt2}) cannot factorise) nor for the third leading jet
% (as Eq.~(\ref{eq:S3pt3}) shows a clear $R$ dependence which cannot simplify).

\subsection{Dependence on $\ptcut$  and $R$: the soft-collinear approximation}
\label{sec:ancoll}
The arguments outlined above may seem somewhat formal.
To gain further analytic insight, it is useful to take a
soft-collinear approximation in which case
Eqs.~(\ref{eq:njetfunct}),(\ref{eq:nthjetfunct}) simplify
considerably. Indeed, if one considers a collinear splitting at a
small angle $\vartheta$, the NLO contribution from a real emission
can be written in simple form by parametrising the final-state momenta as
\begin{equation}
  \label{eq:parCS}
  p_1^\mu=\tilde{p}_a^\mu+\mathcal{O}(k_\perp^2),\quad
  p_2^\mu = (1-z) \tilde{p}_b^\mu + k_\perp^\mu +\mathcal{O}(k_\perp^2), \quad
  p_3^\mu = z \,\tilde{p}_b^\mu - k_\perp^\mu +\mathcal{O}(k_\perp^2),
\end{equation}
where $\tilde{p}_a^\mu$ and $\tilde{p}_b^\mu$ are the Born final-state
hard directions, $z$ is the longitudinal momentum fraction of the
splitting, and the transverse momentum $k_\perp$ satisfies
$k_\perp \cdot \tilde{p}_a = k_\perp \cdot \tilde{p}_b = 0$; $k_\perp$
can then be parametrized by the angle $\vartheta$ between $p_2$ and
$p_3$ and an azimuthal angle $\varphi$.

Including only terms
that produce a logarithmic enhancement in the limit
$\vartheta\to 0$, the real emission contribution takes the form
\begin{equation}\label{eq:colxs}
  d{\Phi}_{3} \frac{d\hat{\sigma}^{(1)}_3}{d {\Phi}_{3} }
  =\sum_{i=q,g} \left[
    \frac{d\sigma^{(0)}_2}{d \ptt} (\pth)
  \right]_i  \left[ \frac{\alpha_s C_i}{\pi} P_i(z) \right]\,
  {\rm d}\pth\, {\rm d}z \frac {{\rm d}\vartheta^2}{\vartheta^2}
  \frac{d \varphi}{2\pi}.
\end{equation}
Note that within this approximation  recoil effects on $p_1$ become
negligible. They  could be addressed using a similar formalism but
going beyond the small-angle approximation that we adopt here.

In Eq.~(\ref{eq:colxs}) $[d\sigma^{(0)}_2/d \ptt]_i$, with $i=q,g$, is the
LO differential cross-section for producing a quark or a gluon of
transverse momentum $\tilde{p}_t$, correctly normalized in such a way
that the sum over $i$ gives the total cross-section. $P_i(z)$
corresponds to the standard Altarelli-Parisi splitting functions with
$z$ the momentum fraction of the collinear splitting (see
Appendix~\ref{app:ancalc} for explicit expressions) from which we have
explicitly factored out a colour factor $2\,C_i$ ($C_i=C_F$ for
quarks and $C_i=C_A$ for gluons). Finally, $\varphi$ is the azimuthal
angle corresponding of the emission with respect to the Born-level parton that
splits.
At this accuracy, the NLO one-loop virtual correction has exactly the same
form as Eq.~(\ref{eq:colxs}) integrated over the full phase-space of
the extra real emission, but with the opposite sign.
In what follows, we further assume that the extra emission is soft so
we can approximate $P_i(z) \approx \frac{1}{z}$.
This soft approximation is made for the sake of simplicity and
can easily be lifted to include the full splitting function.

The soft-collinear approximation is sufficient to obtain results in
fair agreement with the full calculation, and specifically
reproduce three important aspects discussed in
Sect.~\ref{sec:numerics}.
First, we can see explicitly how the cancellation of the $\ptcut$
dependence which happens in the standard case is spoiled for the weighted
definition (A) and restored with definitions (B) and (C).
Second, we are able to identify the $R$ dependence of the second
leading jet with out-of-cone radiation.
Third, we can further study the impact of weighted definitions at
large $\ptt$.
Conversely, working in a soft-collinear approximation, we are
neglecting all recoil effects. This means in particular that the
calculation below will not reproduce the large $K_1$ factor for the
leading jet.
The text below outlines the structure of the calculation and our main
results, deferring additional details to Appendix~\ref{app:ancalc}.

The fact that the real and virtual contributions have the opposite
sign implies that the $N$-jet contribution Eq.~(\ref{eq:njetfunct})
and the $n$-th jet contribution Eq.~(\ref{eq:nthjetfunct}) take
respectively the simple form
\begin{align}
  \label{eq:master-eq-G}
  \frac{d \sigma^{(k)}_{N\,\text{jets}}}{d \ptt}
  &\approx
  \sum_{i=q,g} \frac{C_i}{\pi} \int
  {\rm d}\pth\, {\rm d}z \frac {{\rm d}\vartheta^2}{\vartheta^2}
  \left[
    \frac{d\sigma^{(0)}_2}{d \ptt} (\pth)
  \right]_i \alpha_s\,P_i(z)\,
    \left\{  G_{3 \to {N\,\text{jets}} } -
    G_{2 \to {N\,\text{jets}} } \right\}\\
  \label{eq:IN}
    & \equiv\sum_{\p = q,g} \left[ \frac{d \sigma^{(0)}_2}{d \ptt} (\ptt) \right]_\p
   \frac{C_{\p}}{\pi} \ln\left(\frac{\Rmax^2}{R^2}\right) I_N
\end{align}
and
\begin{align}
  \label{eq:master-eq-S}
  \frac{d \sigma^{(k)}_{n\,\text{-th jet}}}{d \ptt}
  &\approx
  \sum_{i=q,g} \frac{C_i}{\pi} \int
  {\rm d}\pth\, {\rm d}z \frac {{\rm d}\vartheta^2}{\vartheta^2}
  \left[
    \frac{d\sigma^{(0)}_2}{d \ptt} (\pth)
  \right]_i \alpha_s\,P_i(z)\,
    \left\{  S_{3 \to {n\,\text{-th jet}} } -
      S_{2 \to {n\,\text{-th jet}} } \right\}
    \\
    \label{eq:JN}
   & \equiv  \sum_{\p = q,g} \left[ \frac{d \sigma^{(0)}_2}{d \ptt} (\ptt) \right]_\p
   \frac{C_{\p}}{\pi} \ln\left(\frac{\Rmax^2}{R^2}\right) J_n\,,
\end{align}
where in both cases $\Rmax$ is the upper limit of the $\vartheta$ integration.
The functions $I_N$ and $J_n$ can be cast in a simple closed analytic
form by writing the LO cross-section as a power law
\begin{equation}\label{eq:Bornpower}
  \left[
    \frac{d\sigma^{(0)}_2}{d \ptt} (\pth)
  \right]_i \sim \frac{1}{\pth^{m_i}},
\end{equation}
where $m_i$ is, in general, different for the quark and gluon
case.  In 
Appendix~\ref{app:ancalc} explicit analytic expressions 
are given for the standard definition, with the general definitions
easily amenable to numerical treatment. 

We can now use Eqs.~(\ref{eq:IN}),(\ref{eq:JN}) to address the issues
mentioned above.
We start by investigating the behavior in the $\ptt\rightarrow \ptcut$
limit and focus on the leading jet.
$J_1$ receives real contributions from $S_{3\to p_{t1}}$,
Eq.~(\ref{eq:S3pt1}), and virtual corrections from $S_{2\to p_{t1}}$,
Eq.~(\ref{eq:S2}).
The latter contribution cancels against the real one in the region
$\Delta R_{23}\equiv \vartheta<R$. Up to power corrections in $z$, we
can set $\kt{2}=(1-z)\kt{1}$ and $\kt{3}=z\kt{1}$. For $\ptt \to
\ptcut$ we can then assume $\kt{3}<\ptcut$ and we are left with two terms:
\begin{equation}
  J_1 \stackrel{\ptt \to \ptcut}{\sim}
  \int_{1- \ptcut/\ptt}^{\ptcut/\ptt} d z\,P(z)\,w^{(1)}(\ptt;\ptt)
  - \int_{1- \ptcut/\ptt}^{\ptcut/\ptt} d z\,P(z)\,w^{(2)}(\ptt;\ptt,\ptt).
\end{equation}
The first term corresponds to $\kt{2}<\ptcut$ while the second term
includes the real emissions with $\kt{2}>\ptcut$ as well as the
remaining virtual corrections.
After integration over $z$, we thus find
\begin{equation}
  \label{eq:I1}
    J_1 =  \log\left(\frac{\ptcut}{\ptt - \ptcut} \right)
          - \omega \log\left(\frac{\ptcut}{\ptt - \ptcut} \right) 
    =
      \begin{cases}
        0 & \,({\text{standard}})\\
        -\frac{1}{2} \log\left( \frac{\ptt - \ptcut}{\ptcut} \right) & \,
        ({\text{weighted (A)}})
      \end{cases}
      \,,
\end{equation}
where $\omega=1$ for the standard definition and $\omega=\tfrac{1}{2}$
for the weighted definition (A), independently of the exponent $r$
which enters the definition of the weights, Eq.~(\ref{eq:stw}). In the
same limit it turns out that $J_2$ and $J_3$ are nonsingular.  This
explains our findings from Sect.~\ref{sec:numerics}: the unitary
definition suffers from a logarithmic divergence close to $\ptcut$
while the standard definition is independent of the value of
$\ptcut$. Furthermore, this behavior (see
Fig.~\ref{fig:ptfr-weighted}), only affects the leading jet, whose
properties are encoded in $J_1$.
Of course it also follows from
Eq.~\eqref{eq:I1} that when $\ptcut \ll \ptt$, corresponding to using
definitions of the weights of type (B), the singular behavior
disappears.
A similar conclusion can be reached for the definition of type (C).

\begin{figure}[t]
  \centering
  \includegraphics[width=0.45\textwidth]{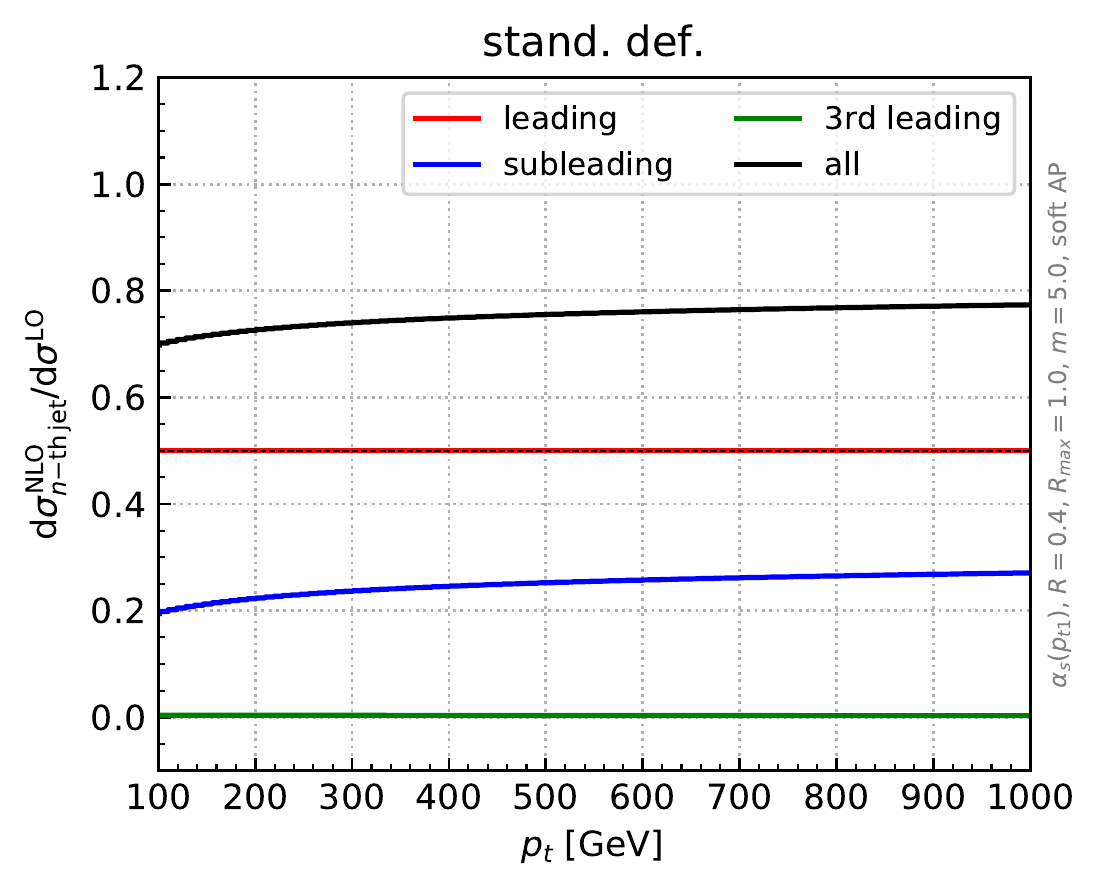}\hfill%
  \includegraphics[width=0.45\textwidth]{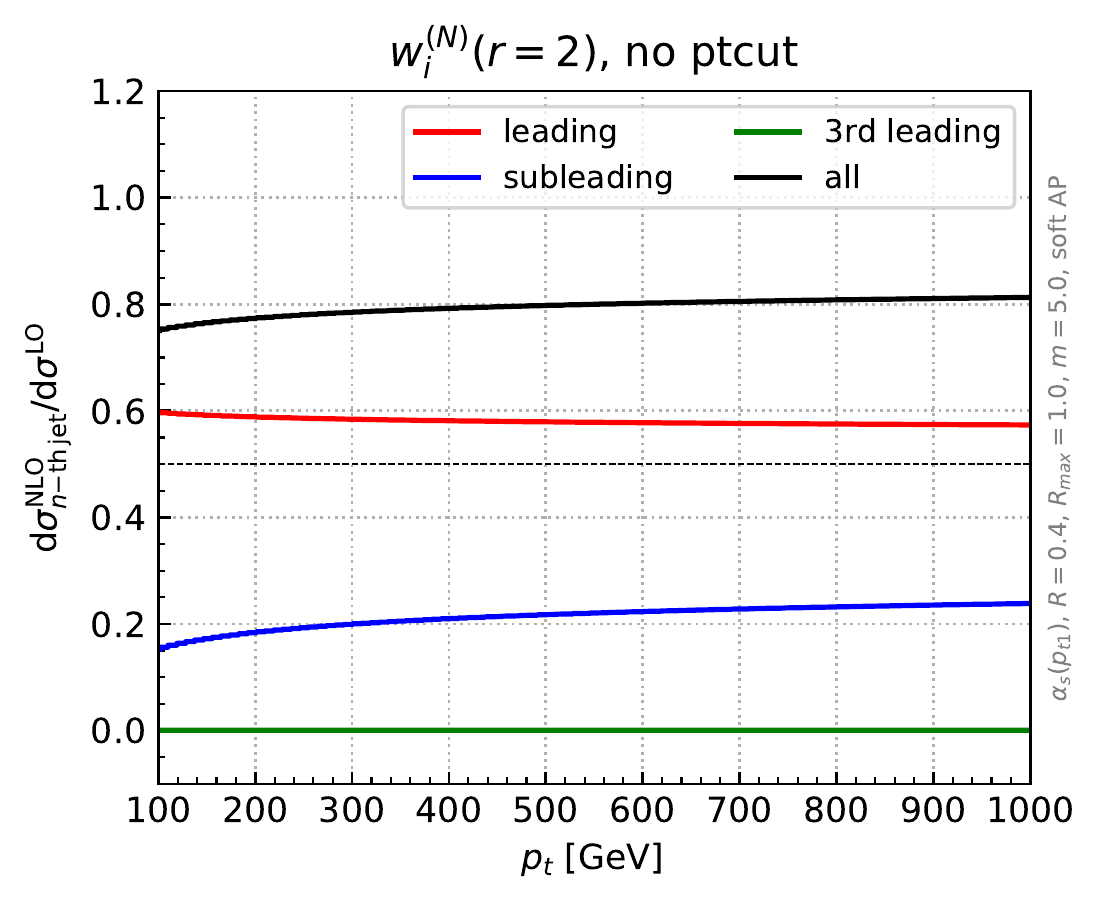} \\
  \includegraphics[width=0.45\textwidth]{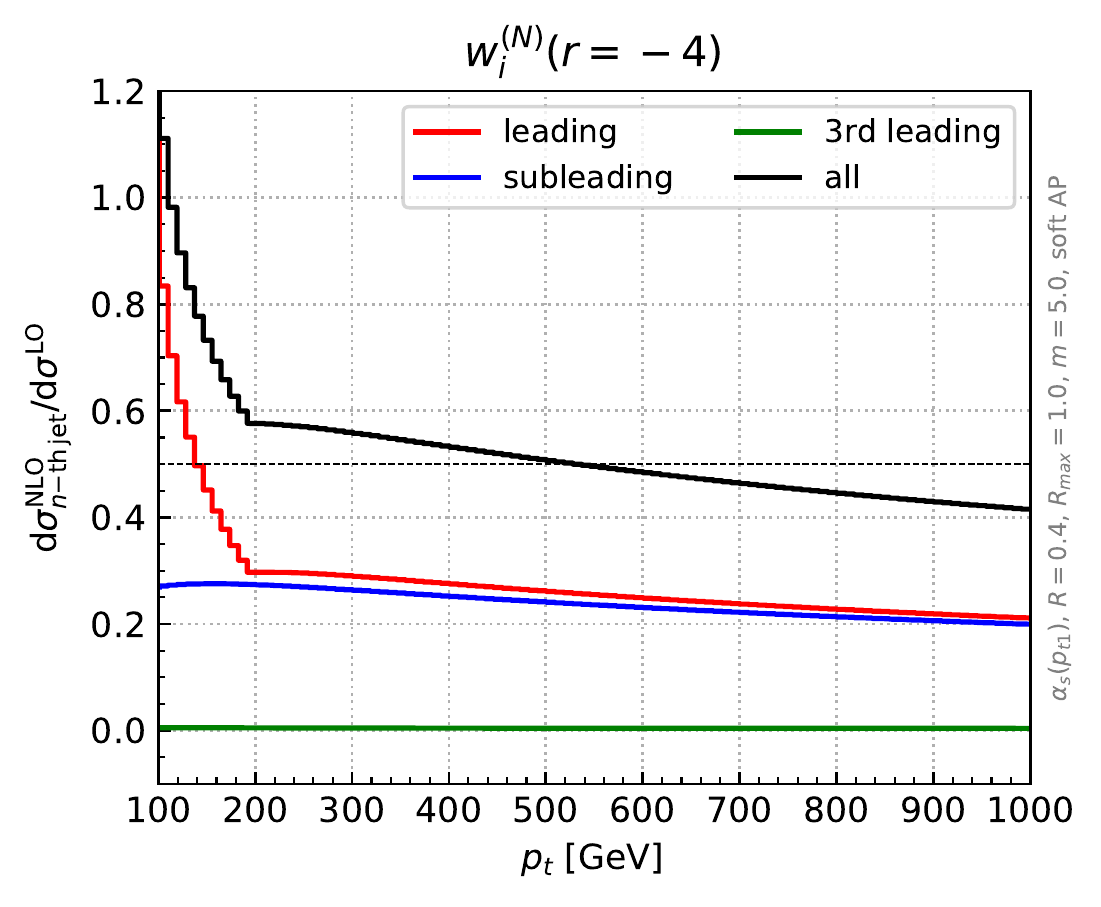}\hfill%
  \includegraphics[width=0.45\textwidth]{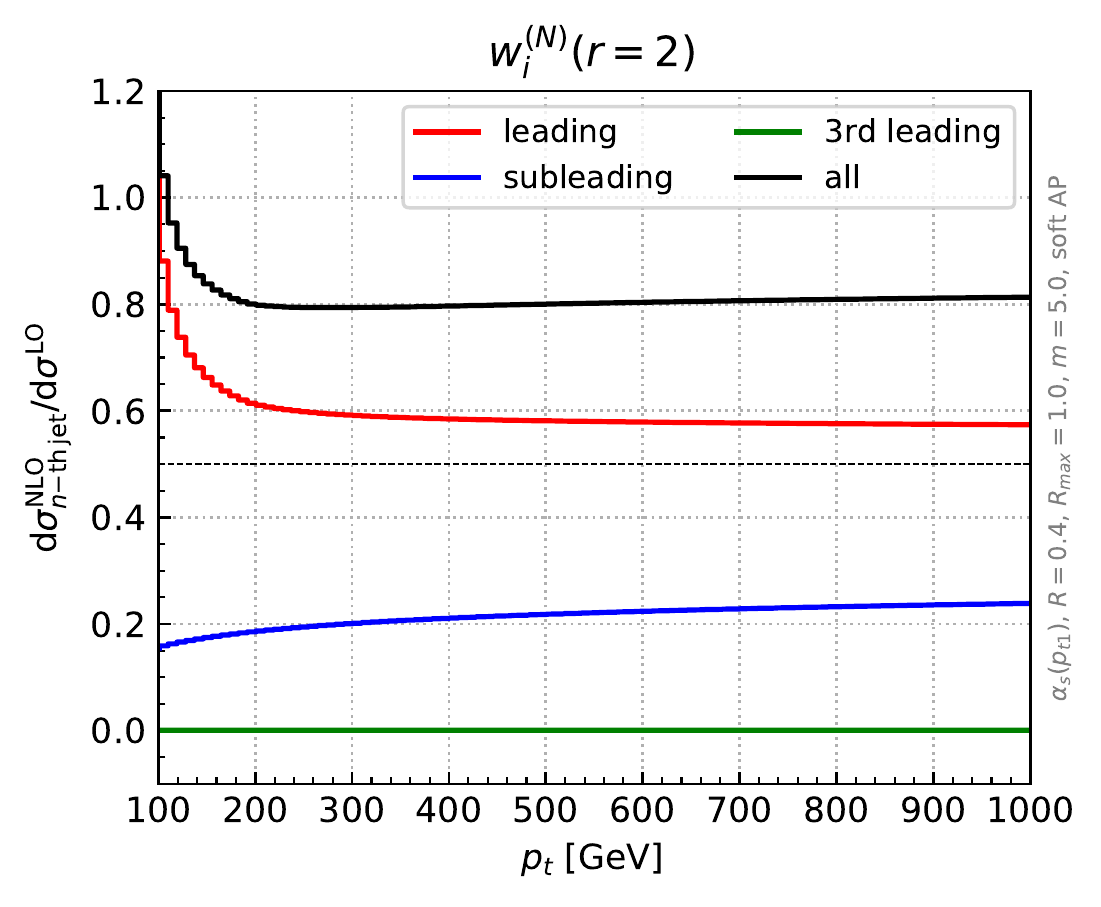} \\ 
  \includegraphics[width=0.45\textwidth]{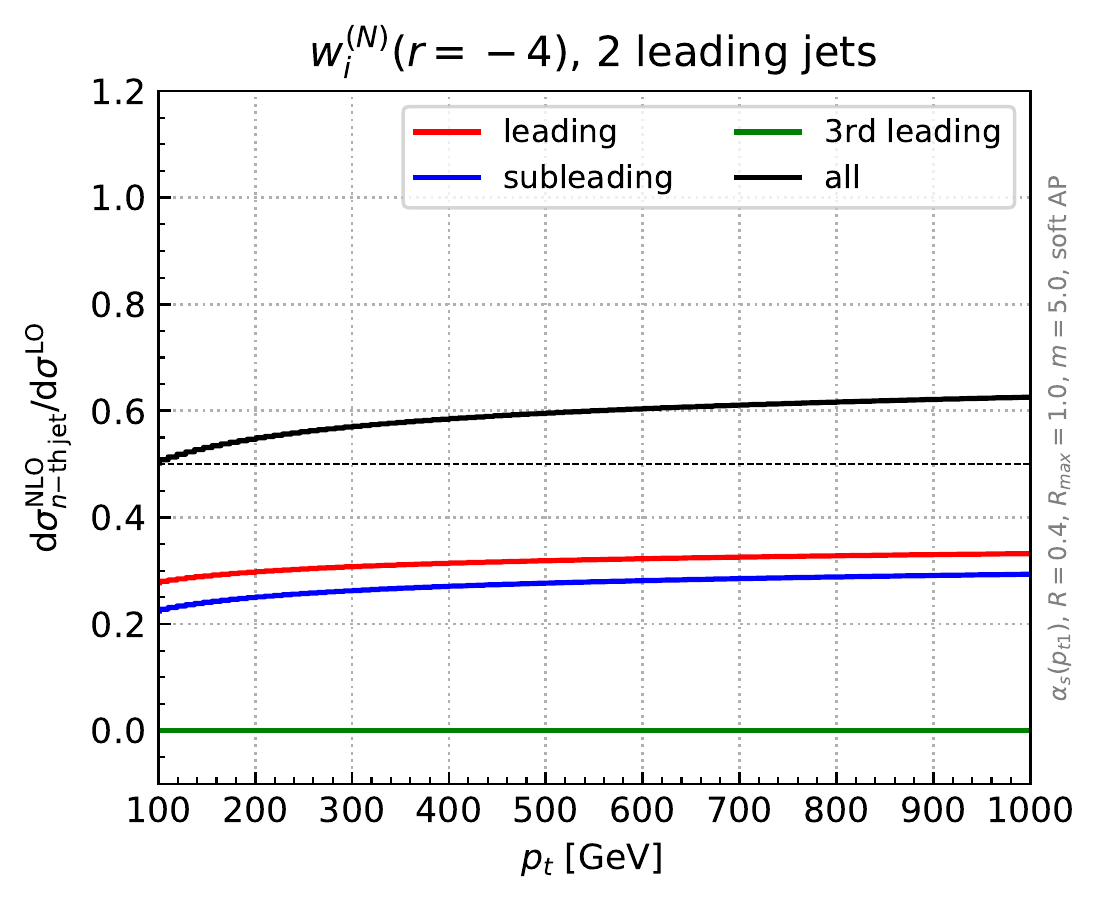}\hfill%
  \includegraphics[width=0.45\textwidth]{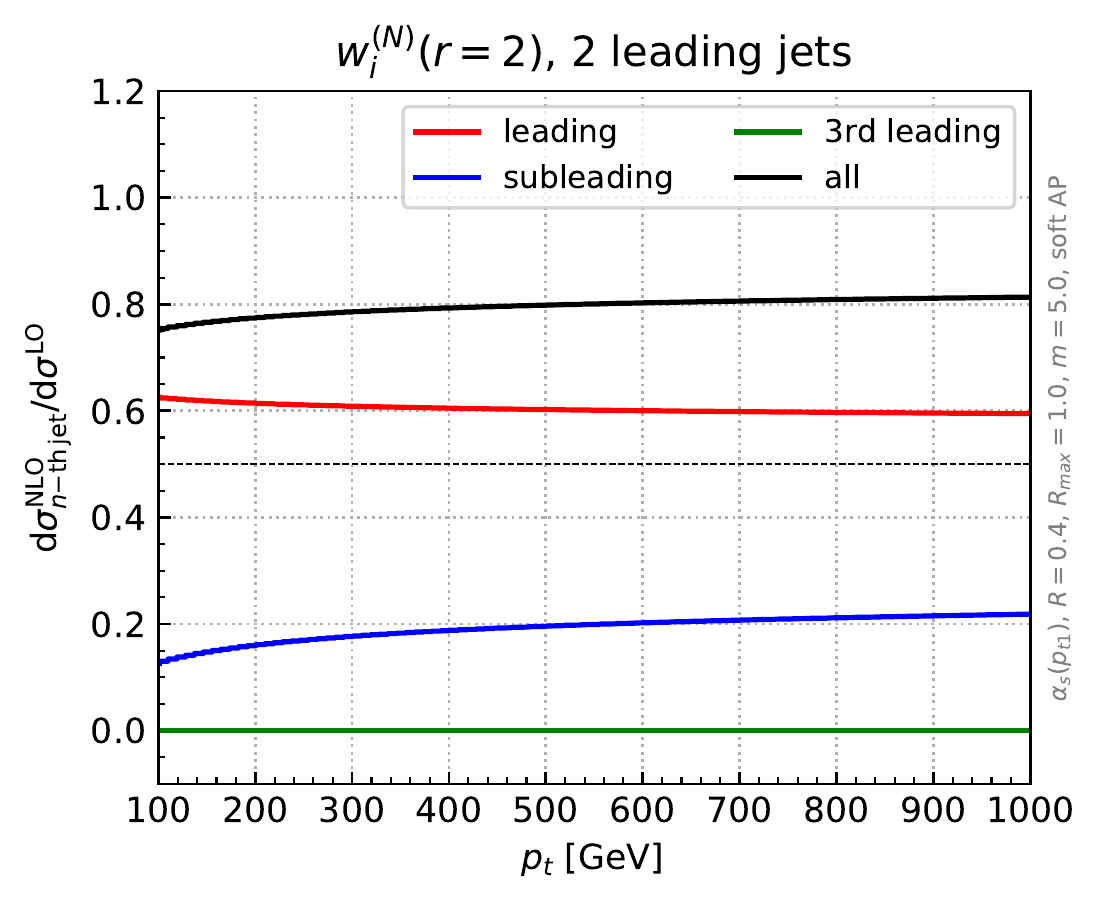} \\
  \caption{Contributions from the leading, subleading, and
    third-leading jets to the NLO inclusive $K$ factors in the
    soft-collinear approximation. The standard definition (top left) is compared
    to weighted definition of type (B) (no $\ptcut$) with $r=2$ (top
    right),
    weighted definitions of type (A) (with $\ptcut)$ with $r=-4$ (middle left) and
    $r=2$ (middle right) and of type (C) (two jets) also with with $r=-4$ (bottom left) and
    $r=2$ (bottom right).}
  \label{fig:analytical}
\end{figure}

Next, we can also use Eq.~(\ref{eq:JN}) to predict the small-$R$
behavior of the second and third leading jet contributions. In both
cases one would get a logarithmic enhancement at small $R$.
Note that at first sight Eq.~(\ref{eq:JN}) seems to imply that the
leading jet contribution also has a logarithmic $R$ dependence in the
standard case, in contradiction to the behavior observed in
Fig.~\ref{fig:KvR}, and to our previous general conclusion based on
Eq.~(\ref{eq:Slj}).
However, one should realise that, in the small-$R$ limit where
Eq.~(\ref{eq:JN}) holds, Eq.~\eqref{eq:I1} implies that $J_1$ is zero,
and thus obviously $R$-independent in the standard case. In all
weighted cases $J_1$ is non-vanishing, and thus the leading jet
contribution becomes $R$-dependent in agreement with our previous
analytic and numerical arguments, with a logarithmic dependence on $R$
in the small-$R$ limit.

Finally, we can study the limit of the functions $J_n$ when
$\ptt\gg\ptcut$, in the weighted case with $r$ negative and
$|r|\sim |m_i| \sim 4$.  In this case, we find that the contributions
from the leading and the subleading jet are comparable (see the
Appendix~\ref{app:ancalc} for details), partially solving the problem
of the large compensation seen in the standard definition or for
positive values of $r$, as observed in Sect.~\ref{sec:numerics-wgt},
Fig.~\ref{fig:ptfr-weighted}.

Results obtained for the leading, subleading and third-leading jet contributions
using the approximation
Eqs.~(\ref{eq:IN}),(\ref{eq:JN}) are shown in Fig.~\ref{fig:analytical}
for a representative set of cases, to be compared to Figs.~\ref{fig:ptfr},\ref{fig:ptfr-weighted}-\ref{fig:ptfr-weighted-2leading}.
All plots have been produced implementing
Eq.~\eqref{eq:master-eq-S}, with Eq.~\eqref{eq:Bornpower} and $m=5$.
Note that this
parametrization of the LO $\ptt$ spectrum already includes initial
state PDFs. We have checked that using the exact LO partonic
cross-section yields similar results. 
We choose $R_{\text{max}}=1$ and use $R=0.4$
to allow for a comparison with the full results presented in Section~\ref{sec:numerics}. 
Finally, we set $\alpha_s(\pt{1})$.
As anticipated, it is clear that the main qualitative features of the exact
results are reproduced by the soft-collinear approximation.

\section{Conclusions}
\label{sec:conclusion}

\begin{table}
  \centering
  \begin{tabular}{|l|c|c|c|c|}
    \hline
    \multirow{2}{*}{Definition}
    & \multirow{2}{*}{standard}
    & \multicolumn{3}{|c|}{weighted} \\
    \cline{3-5}
    & 
    & (A) above $\ptcut$
    & (B) all jets
    & (C) two leading\\
    \hline
    Reference plot
    & Fig.~\ref{fig:ptfr}
    & Fig.~\ref{fig:ptfr-weighted}
    & Fig.~\ref{fig:ptfr-weighted-all}
    & Fig.~\ref{fig:ptfr-weighted-2leading}\\
    \hline
    \hline
    unitarity & no & yes & yes & yes \\
    \hline
    no large logs
    & \multirow{2}{*}{$\checkmark$}
    & \multirow{2}{*}{\xmark}
    & \multirow{2}{*}{$\checkmark$}
    & \multirow{2}{*}{$\checkmark$} \\
    close to $\ptcut$ &&&&\\
    \hline
    no large logs
    & \multirow{2}{*}{$\checkmark$}
    & $\checkmark$ for $r>0$
    & \multirow{2}{*}{$\checkmark$}
    & \multirow{2}{*}{$\checkmark$} \\
    at large $p_t$
    & & \xmark\ for $r<0$ & & \\
    \hline
    overlapping scale 
    & \xmark
    & \multirow{2}{*}{$\checkmark$}
    & \multirow{2}{*}{$\checkmark$}
    & \multirow{2}{*}{$\checkmark$}\\
    variation bands
    & $\checkmark$ with uncorr. uncert.~\cite{Dasgupta:2016bnd,Currie:2016bfm}
    & & & \\
    \hline
    no large cancellations
    & \multirow{2}{*}{\xmark}
    & \multirow{2}{*}{\xmark}
    & \multirow{2}{*}{\xmark}
    & \xmark\ for $r>0$\\
    between $K_1$ and $K_2$      
    & & & & $\checkmark$ for $r<0$\\
    \hline
  \end{tabular}
  \caption{Summary of the main properties of the various
    single-inclusive jet definitions studied in this
    paper.}\label{tab:summary}
\end{table}

In this paper we have addressed the potential issue of the
non-unitarity of the single-jet inclusive cross-section, by
introducing a series of alternative weighted definitions of this
observable which are unitary in the sense that upon integration they
lead to the total cross-section.
The main features of the various definitions we have considered are
summarised in Table~\ref{tab:summary}.

Our conclusion is that a  naive weighted approach [type (A) of
  Sect.~\ref{sec:definition}] in which
one simply introduces a weighting of all jets above a certain $\ptcut$
is flawed, in the sense that it develops logarithmic singularities
associated with the transverse momentum cut on jets, $\ptcut$.
More sophisticated definitions avoid this problem by setting $\ptcut$
to zero [type (B)] or by considering only the two leading jets [type
(C)]. Both these definitions could be more challenging to implement in
a practical (experimental) environment.

Additionally, even leaving aside practical considerations, there does
not seems to be any real advantage in adopting these definitions in
term of perturbative stability. In particular, all weighted
definitions with positive $r$ show features at best similar to the
standard definition.
Furthermore, the apparent perturbative instability of the conventional
definition appears in fact to be the manifestation of an unnatural
smallness of the NLO $K$ factors which only happens for a limited
range of jet radius $R\sim 0.4$. It is a consequence of an accidental
cancellation which makes standard scale variation unreliable as a
means of estimating missing higher order corrections. This apparent
issue for example disappears with more conservative estimates of the
perturbative uncertainties.
One possible case of interest is the definition of type (C), focusing
on the two leading jets, with $r<0$. Compared to the standard
definition, it has the potential advantage of reducing the large
difference between the $K$ factor of the leading and subleading jets,
at the cost of having a larger overall NLO $K$ factor. 

Our final conclusion is both negative, and positive. On the negative side,
we conclude
that unitary definitions of the jet inclusive
cross-section are at best as good as the standard definition, while
being rather more contrived. On the positive side, we conclude that
the standard definition  shows no critical sign of pathological
features or problems, other  than its
unitarity, which however is per se not causing any perturbative
problem. Among the unitary definitions, the weighted definitions
based on including only the two
leading jets appear to be particularly well-behaved.
While in this work we study the dijet system as a 
function of the $p_t$ and rapidity of the individual jets, this is in 
agreement  with previous studies~\cite{Currie:2018xkj} in which dijet
observables are also  
found to have better perturbative stability.

\acknowledgments

We thank Jesse Thaler for a number of interesting discussions
at various times during the completion of this work.
Stefano Forte is supported by the European
Research Council under the European Union's Horizon
2020 research and innovation Programme (grant agreement n.740006).
Matteo Cacciari, Davide Napoletano and Gregory Soyez are supported in part by the
French Agence Nationale de la Recherche, under grant ANR-15-CE31-0016.

\appendix
%\section{Jet Functions}
%\label{app:jetfunc}
\section{NLO cross-section in the soft-collinear approximation}
\label{app:ancalc}

The $N$-jet contribution and the $n$-th jet contribution to the differential
cross-section at NLO in the soft-collinear approximation are given by
Eq.~(\ref{eq:master-eq-G}) and Eq.~(\ref{eq:master-eq-S}) respectively.
Using an explicit expression for the splitting
functions $P_i$ and for the $G$ or the $S$ functions in the collinear
limit we can  perform the phase-space integration explicitly.

The splitting functions $P_i$ are
\begin{equation}
  \begin{split}\label{eq:splita}
    P_q(z) &= \sum_{j=q,g}P_{jq} = \frac{1+(1-z)^2}{2z} = \frac{1}{z} + \mathcal{O}(1) \\
    P_g(z) &= \sum_{j=q,g}P_{jg} = \frac{1}{2} \left[
      2 \frac{1-z}{z} + z(1-z) + \frac{T_R N_f}{C_A} (z^2 + (1-z)^2)
    \right] = \frac{1}{z} + \mathcal{O}(1)
  \end{split}
\end{equation}
where the $z \leftrightarrow (1-z)$ symmetry has been exploited in such a way
that all soft-collinear singularities are at $z=0$ (see e.g.~\cite{jetbook}).
Note that a $2C_F$ or $2 C_A$ factor, respectively, has been explicitly
factored out.

By adopting the parametrization of the final-state given
in~Eq.~(\ref{eq:parCS}), the jet functions $G$ and $S$ can be rewritten in the
collinear and small $R$  limit, i.e. $\Delta R_{23} = \vartheta \leq R \ll 1$.
For the weighted definition with jets above $\ptcut$ we have:
\begingroup
\allowdisplaybreaks
\begin{align}
  \label{eq:jetf-v-smallR}
  G_{2 \to 1} & = G_{2 \to 3} = 0\\
  \label{eq:jetf-v2-smallR}
  G_{2 \to 2} &= \Theta(\pth > \ptcut) w^{(2)}(\ptt|\pth, \pth) 
  [\delta(\ptt - \pth) + \delta(\ptt - \pth)]\\
  \label{eq:jetf1-smallR}
  G_{3 \to 1} & = \Theta(\vartheta^2 > R^2)
     \Theta(\pth > \ptcut; z \pth < \ptcut; (1-z)\pth < \ptcut)
     w^{(1)}(\ptt|\pth) [\delta(\ptt - \pth)]  \\
  G_{3 \to 2} & = \Theta(\vartheta^2 < R^2)
    \Theta(\pth > \ptcut) w^{(2)}(\ptt|\pth, \pth) 
    [\delta(\ptt - \pth) + \delta(\ptt - \pth)] \nonumber \\
  & + \Theta(\vartheta^2 > R^2) \Theta(\pth > \ptcut)\nonumber\\
  & \phantom{+} \Big\{\Theta(z \pth < \ptcut; (1-z)\pth > \ptcut)
    w^{(2)}(\ptt|\pth, (1-z)\pth)  [\delta(\ptt - \pth) + \delta(\ptt - (1-z)\pth)]\nonumber \\
  & \phantom{+}+ \Theta(z \pth > \ptcut; (1-z)\pth < \ptcut)
    w^{(2)}(\ptt|\pth, z\pth)
    [\delta(\ptt - \pth) + \delta(\ptt - z\pth)] \Big\} \label{eq:jetf2-smallR}\\
  \label{eq:jetf3-smallR}
  G_{3 \to 3} & = \Theta(\vartheta^2 > R^2)
    \Theta(\pth > \ptcut; z \pth > \ptcut; (1-z)\pth > \ptcut)
     \nonumber \\ 
    &\phantom{ ==}  w^{(3)}(\ptt|\pth, z\pth, (1-z)\pth)
      [\delta(\ptt - \pth) + \delta(\ptt - z\pth) + \delta(\ptt - (1-z)\pth)].
\end{align}
\endgroup
and
\begingroup
\allowdisplaybreaks
\begin{align}
  \label{eq:S2-smallR}
  S_{2 \to \pt{1}}
  & =  S_{2 \to \pt{2}}
    = \Theta(\pth > \ptcut)  w^{(2)}(\ptt|\pth, \pth) \delta(\ptt - \pth)\\
  \label{eq:S3pt1-smallR}
  S_{3 \to \pt{1}}
  &= \Theta(\vartheta^2 < R^2)
    \Theta(\pth > \ptcut)  w^{(2)}(\ptt|\pth, \pth)  \delta(\ptt -  \pth)
    + \Theta(\vartheta^2 > R^2)  \\
  &\quad  \Big[ \Theta(\pth > \ptcut; z \pth < \ptcut; (1-z)\pth < \ptcut)
    w^{(1)}(\ptt|\pth) \delta(\ptt - \pth) \nonumber\\
  &\quad + \Theta(\pth > \ptcut; z \pth > \ptcut; (1-z)\pth < \ptcut)
    w^{(2)}(\ptt|\pth, z \pth) \delta(\ptt - \pth)  \nonumber\\
  &\quad + \Theta(\pth > \ptcut; z \pth < \ptcut; (1-z)\pth > \ptcut)
    w^{(2)}(\ptt|\pth, (1-z)\pth) \delta(\ptt - \pth) \nonumber\\
  &\quad + \Theta(\pth > \ptcut; z \pth > \ptcut; (1-z)\pth > \ptcut)
    w^{(3)}(\ptt|\pth, z \pth, (1-z)\pth) \delta(\ptt - \pth) \Big]\nonumber\\
  \label{eq:S3pt2-smallR}
  S_{3 \to \pt{2}}
  &= \Theta(\vartheta^2 < R^2) 
      \Theta(\pth > \ptcut)  w^{(2)}(\ptt|\pth, \pth)  \delta(\ptt - \pth)
     + \Theta(\vartheta^2 > R^2) \\
  &\quad  \Big\{  \Theta(\pth > \ptcut; z \pth > \ptcut; (1-z)\pth < \ptcut)
    w^{(2)}(\ptt|\pth, z \pth) \delta(\ptt - z\pth) \nonumber\\
  &\quad + \Theta(\pth > \ptcut; z \pth < \ptcut; (1-z)\pth > \ptcut)
    w^{(2)}(\ptt|\pth, (1-z)\pth) \delta(\ptt - (1-z)\pth) \nonumber\\
  &\quad + \Theta(\pth > \ptcut; z \pth > \ptcut; (1-z)\pth > \ptcut)
    w^{(3)}(\ptt|\pth, z \pth, (1-z)\pth) \nonumber\\
  &\quad\quad\quad\quad   \left[
    \Theta(z > 1/2) \delta(\ptt - z\pth) + \Theta(z < 1/2) \delta(\ptt - (1-z)\pth)
    \right] \Big\}\nonumber\\
  S_{3 \to \pt{3}}
  & = \Theta(\vartheta^2 > R^2) 
    \Theta(\pth > \ptcut; z \pth > \ptcut; (1-z)\pth > \ptcut)
    w^{(3)}(\ptt|\pth, z \pth, (1-z)\pth)\nonumber\\
  & \phantom{=}  \left[
    \Theta(z < 1/2) \delta(\ptt - z\pth) + \Theta(z > 1/2) \delta(\ptt - (1-z)\pth)
    \right]  \label{eq:S3pt3-smallR}
\end{align}
\endgroup The standard definition can trivially be recovered by
setting the weights to 1, while the case of the weighted definition
including all jets can be obtained by taking the limit $\ptcut\to 0$.
Similarly, the weighted definition with 2 leading jets is instead
obtained by firstly taking the limit $\ptcut\to 0$ and by then keeping
the terms proportional to $\delta(\ptt - \pth)$ as well as the terms
proportional to either $\delta(\ptt - z\pth)$ if $z > 1/2$, or
$\delta(\ptt - (1-z)\pth)$ if $z < 1/2$, modifying the weights
accordingly.

The $\pth$ integration in
Eqs.~(\ref{eq:master-eq-G})-(\ref{eq:master-eq-S}) can be simplified
using the delta functions $\delta(\ptt - \pth)$,
$\delta(\ptt - z\pth)$ and $\delta(\ptt - (1-z)\pth)$. The $\vartheta$
integration leads to a logarithmic dependence on the jet radius
$R$. The only nontrivial integral is over $z$, thereby leading to a
final result of the form of Eqs.~(\ref{eq:IN}),(\ref{eq:JN}).
Explicitly, $I_N$ and $J_n$ there present are given by:
\begin{equation}
  \label{eq:I1expl}
  I_1 =
  \Theta(\ptt < 2\ptcut) \int_{1- \ptcut/\ptt}^{\ptcut/\ptt} d z\,P(z) 
  \left[1\right] \sigmatilde(\ptt)
\end{equation}
\begin{equation}
  \begin{split}
    \label{eq:I2expl}
    I_2 =&
    \Theta(\ptt < 2\ptcut)
    \Bigg[
    \int_{\ptcut/\ptt}^{1} d z\,P(z)
    \left[ \frac{1}{1+z^r} \right] \sigmatilde(\ptt) \\
    &\phantom{=} +
    \int_0^{1- \ptcut/\ptt} d z\,P(z)
    \left(
      \left[ \frac{1}{1+(1-z)^r} \right] \sigmatilde(\ptt) - \left[ \frac{1}{2} \right] \sigmatilde(\ptt)
    \right) 
    - \int_{1- \ptcut/\ptt}^1 d z\,P(z) \left[ \frac{1}{2} \right] \sigmatilde(\ptt)
    \Bigg] \\
    &+
        \Theta(\ptt > 2\ptcut)
    \Bigg[
    \int_{1- \ptcut/\ptt}^{1} d z\,P(z)
    \left[ \frac{1}{1+z^r} \right] \sigmatilde(\ptt) \\
    &\phantom{=} +
    \int_0^{\ptcut/\ptt} d z\,P(z)
    \left(
      \left[ \frac{1}{1+(1-z)^r} \right] \sigmatilde(\ptt) - \left[ \frac{1}{2} \right] \sigmatilde(\ptt)
    \right)
    - \int_{\ptcut/\ptt}^1 d z\,P(z) \left[ \frac{1}{2} \right] \sigmatilde(\ptt)
    \Bigg] \\
    &+
    \int_{\ptt/(\ptt + \ptcut)}^{1} d z\,P(z)
    \left[ \frac{z^r}{1+z^r} \right] \frac{1}{z} \sigmatilde\left(\frac{\ptt}{z}\right) \\
    &+
    \int_0^{\ptcut/(\ptt + \ptcut)} d z\,P(z)
    \left(
      \left[ \frac{(1-z)^r}{1+(1-z)^r} \right] \frac{1}{1-z} \sigmatilde\left(\frac{\ptt}{1-z}\right)
      - \left[ \frac{1}{2} \right] \sigmatilde(\ptt)     
    \right)  \\
    &- \int_{\ptcut/(\ptt + \ptcut)}^1 d z\,P(z) \left[ \frac{1}{2} \right] \sigmatilde(\ptt)
  \end{split}
\end{equation}
\begin{equation}
  \begin{split}
    \label{eq:I3expl}
    I_3 =&
    \Theta(\ptt > 2\ptcut) \int_{\ptcut/\ptt}^{1-\ptcut/\ptt} d z\,P(z) 
    \left[\frac{1}{1+z^r+(1-z)^r}\right] \sigmatilde(\ptt) \\
    &+
    \int_{0}^{\ptt/(\ptt + \ptcut)} d z\,P(z) 
    \left[\frac{z^r}{1+z^r+(1-z)^r}\right] \frac{1}{z} \sigmatilde\left(\frac{\ptt}{z}\right) \\
    &+
    \int_{\ptcut/(\ptt + \ptcut)}^{1} d z\,P(z) 
    \left[\frac{(1-z)^r}{1+z^r+(1-z)^r}\right] \frac{1}{1-z} \sigmatilde\left(\frac{\ptt}{1-z}\right) \\
  \end{split}
\end{equation}
\begin{equation}
  \begin{split}
    \label{eq:J1expl}
    J_1 =&
    \Theta(\ptt < 2\ptcut)
    \Bigg[
    \int_{1- \ptcut/\ptt}^{\ptcut/\ptt} d z\,P(z) 
    \left[1\right] \sigmatilde(\ptt)
    - \int_{1- \ptcut/\ptt}^1 d z\,P(z) \left[ \frac{1}{2} \right] \sigmatilde(\ptt) \\
    &\phantom{=} +
    \int_0^{1- \ptcut/\ptt} d z\,P(z)
    \left(
      \left[ \frac{1}{1+(1-z)^r} \right] \sigmatilde(\ptt) - \left[ \frac{1}{2} \right] \sigmatilde(\ptt)
    \right)
    +
    \int_{\ptcut/\ptt}^{1} d z\,P(z)
    \left[ \frac{1}{1+z^r} \right] \sigmatilde(\ptt) 
    \Bigg] \\
    &+
    \Theta(\ptt > 2\ptcut)
    \Bigg[
    \int_{\ptcut/\ptt}^{1-\ptcut/\ptt} d z\,P(z) 
    \left[\frac{1}{1+z^r+(1-z)^r}\right] \sigmatilde(\ptt)
    - \int_{\ptcut/\ptt}^1 d z\,P(z) \left[ \frac{1}{2} \right] \sigmatilde(\ptt) \\
    &\phantom{=} +
    \int_0^{\ptcut/\ptt} d z\,P(z)
    \left(
      \left[ \frac{1}{1+(1-z)^r} \right] \sigmatilde(\ptt) - \left[ \frac{1}{2} \right] \sigmatilde(\ptt)
    \right)
    +
    \int_{1- \ptcut/\ptt}^{1} d z\,P(z)
    \left[ \frac{1}{1+z^r} \right] \sigmatilde(\ptt)
    \Bigg] \\
  \end{split}
\end{equation}
\begin{equation}
  \begin{split}
    \label{eq:J2expl}
    J_2 =&
    \int_{\ptt/(\ptt + \ptcut)}^{1} d z\,P(z)
    \left[ \frac{z^r}{1+z^r} \right] \frac{1}{z} \sigmatilde\left(\frac{\ptt}{z}\right)
    - \int_{\ptcut/(\ptt + \ptcut)}^1 d z\,P(z) \left[ \frac{1}{2} \right] \sigmatilde(\ptt) \\
    &+
    \int_0^{\ptcut/(\ptt + \ptcut)} d z\,P(z)
    \left(
      \left[ \frac{(1-z)^r}{1+(1-z)^r} \right] \frac{1}{1-z} \sigmatilde\left(\frac{\ptt}{1-z}\right)
      - \left[ \frac{1}{2} \right] \sigmatilde(\ptt)     
    \right) \\
    &+
    \int_{1/2}^{\ptt/(\ptt + \ptcut)} d z\,P(z) 
    \left[\frac{z^r}{1+z^r+(1-z)^r}\right] \frac{1}{z} \sigmatilde\left(\frac{\ptt}{z}\right) \\
    &+
    \int_{\ptcut/(\ptt + \ptcut)}^{1/2} d z\,P(z) 
    \left[\frac{(1-z)^r}{1+z^r+(1-z)^r}\right] \frac{1}{1-z} \sigmatilde\left(\frac{\ptt}{1-z}\right)  
  \end{split}
\end{equation}
\begin{equation}
  \begin{split}
    \label{eq:J3expl}
    J_3 =&
    \int_{0}^{1/2} d z\,P(z) 
    \left[\frac{z^r}{1+z^r+(1-z)^r}\right] \frac{1}{z} \sigmatilde\left(\frac{\ptt}{z}\right)
    +
    \int_{1/2}^{1} d z\,P(z) 
    \left[\frac{(1-z)^r}{1+z^r+(1-z)^r}\right] \frac{1}{1-z} \sigmatilde\left(\frac{\ptt}{1-z}\right)
  \end{split}
\end{equation}
where the terms in squared brackets correspond to the weights, here
given for a definition of type (A), and we have set the running
coupling scale to $p_t^{\text{max}}\equiv p_{t1}=\tilde{p}_t$ and
introduced
\begin{equation}
  \sigmatilde(x) \equiv \frac{d\sigma^{(0)}_2}{d \ptt} (x) \, \alpha_s(x).
\end{equation}
In the fixed coupling approximation or if we take $\alpha_s(\ptt)$,
the coupling can be factorized out of the integration and directly moved to
Eq.~(\ref{eq:IN}) or Eq.~(\ref{eq:JN}).
Note that the above expressions do not assume $z \ll 1$. Keeping the
full $z$ dependence of the splitting functions would
therefore account for hard-collinear splittings.

In the general weighted case, these integrals can only be computed numerically.
Results for the standard (unweighted) definition are found by simply
removing all terms in square brackets.  In this case, by using
Eq.~(\ref{eq:Bornpower}) for the Born cross-section and the soft
approximation of the splitting functions these
integrals can be computed exactly in the fixed coupling approximation
and their expressions are
\begin{align}
  \label{eq:I1std}
  I_1^\text{(std)} & = \Theta(\ptt < 2\ptcut)\,\ln \left( \frac{\ptcut}{\ptt - \ptcut} \right)\\
  \label{eq:I2std}
    I_2^\text{(std)} &=
    \Theta(\ptt > 2\ptcut)\,\ln \left( \frac{\ptcut}{\ptt - \ptcut} \right) 
    -
    \Theta(\ptt < 2\ptcut)\,\ln \left( \frac{\ptcut}{\ptt - \ptcut} \right)
    + \ln \left( \frac{\ptcut}{\ptt + \ptcut} \right) \\
    &\phantom{=}
    + \frac{1}{m-1} \left( 1 - \left( \frac{\ptt + \ptcut}{\ptt} \right)^{1-m} \right)
    -(m-1) \left( \frac{\ptcut}{\ptt + \ptcut} \right)
    \, _3F_2\left(1,1,2-m; 2,2; \frac{\ptcut}{\ptt + \ptcut} \right)\nonumber\\
    \label{eq:I3std}
      I_3^\text{(std)} &=
      - \Theta(\ptt > 2\ptcut)\,\ln \left( \frac{\ptcut}{\ptt - \ptcut} \right)
      - \ln \left( \frac{\ptcut}{\ptt + \ptcut} \right) \\
      &\phantom{=}
      + \frac{1}{m-1} \left( \frac{\ptt + \ptcut}{\ptt} \right)^{1-m}
      - H_{m-1} + (m-1) \left( \frac{\ptcut}{\ptt + \ptcut} \right)
      \, _3F_2\left(1,1,2-m; 2,2; \frac{\ptcut}{\ptt + \ptcut} \right) \nonumber
\end{align}
and
\begin{align}
    \label{eq:J1std}
    J_1^\text{(std)} & = 0\\
    \label{eq:J2std}
    J_2^\text{(std)} & = -\frac{1}{2} (m-1) \,
    _3F_2\left(1,1,2-m;2,2;\frac{1}{2}\right)
    -\frac{2^{1-m}-1}{m-1} - \log 2\\
    \label{eq:J3std}
    J_3^\text{(std)} & = \frac{1}{2} (m-1) \, _3F_2\left(1,1,2-m;2,2;\frac{1}{2}\right)
    -H_{m-1}+\frac{2^{1-m}}{m-1}+\log 2
\end{align}
where $H_n$ are harmonic numbers, ${}_pF_q$ is a generalized hypergeometric
function, and $m$ is the power of the LO cross-section in
Eq.~(\ref{eq:Bornpower}), which can in principle differ for quarks and gluons.

Adding up all contributions we get:
\begin{equation}
  \begin{split}
  \frac{d \sigma^{(k)}}{d \ptt} &= 
  \sum_{\p = q,g} \left[ \frac{d \sigma^{(0)}_2}{d \ptt} (\ptt) \right]_\p
  \frac{\alpha_s C_{\p}}{\pi} \ln\left(\frac{\Rmax^2}{R^2}\right)
  (I_1 + I_2 + I_3)^\text{(std)} \\
  &=  \sum_{\p = q,g} \left[ \frac{d \sigma^{(0)}_2}{d \ptt} (\ptt) \right]_\p
  \frac{\alpha_s C_{\p}}{\pi} \ln\left(\frac{\Rmax^2}{R^2}\right)
  (J_1 + J_2 + J_3)^\text{(std)} \\
  &= \sum_{\p = q,g} \left[ \frac{d \sigma^{(0)}_2}{d \ptt} (\ptt) \right]_\p
  \frac{\alpha_s C_{\p}}{\pi} \ln\left(\frac{\Rmax^2}{R^2}\right)
  \left[
    \frac{1}{m_\p-1} - H_{m_\p-1}
  \right].
\end{split}
\end{equation}
% i.e. the $K$ factor is flat, since both the $\ptt$ and the $\ptcut$
% dependence have canceled completely in the square bracket in the last
% line.  The only remaining dependence on $\ptt$ can come from the
% running of $\alpha_s$, here neglected.
% 
For $m_q=m_g$,  the $K$ factor is flat, since both the $\ptt$ and the $\ptcut$
dependence have canceled completely in the square bracket in the last
line.  The only remaining dependence on $\ptt$ would therefore come
either from differences between the quark and gluon contributions
($m_q\neq m_g$) or from the running of $\alpha_s$ which was neglected
in the above result.

We conclude by studying the large $\ptt$ limit of $J_n$ in the
weighted case. When $\ptt\to\infty$, from Eqs.~(\ref{eq:J1expl})-(\ref{eq:J2expl})
we get
\begin{align}
  \label{eq:pttoinf}
  J_1^\text{(wgt)} \stackrel{\ptt \to \infty}{\sim} & 
  - \int_{0}^1 d z\,P(z) \left[ \frac{1}{2} \right] \sigmatilde(\ptt)
  + \int_{0}^{1} d z\,P(z) \left[\frac{1}{1+z^r+(1-z)^r}\right] \sigmatilde(\ptt) \\
  J_2^\text{(wgt)} \stackrel{\ptt \to \infty}{\sim} & 
    - \int_{0}^1 d z\,P(z) \left[ \frac{1}{2} \right] \sigmatilde(\ptt) \nonumber\\
    & + \int_{1/2}^{1} d z\,P(z) 
    \left[\frac{z^r}{1+z^r+(1-z)^r}\right] \frac{1}{z} \sigmatilde\left(\frac{\ptt}{z}\right) \nonumber\\
    & + \int_{0}^{1/2} d z\,P(z) 
  \left[\frac{(1-z)^r}{1+z^r+(1-z)^r}\right] \frac{1}{1-z} \sigmatilde\left(\frac{\ptt}{1-z}\right)
\end{align}
while $J_3$ in Eq.~(\ref{eq:J3expl}) does not depend on $\ptt$ and it is always
negligible.  Assuming that the LO cross-section behaves
accordingly to the power law Eq.~(\ref{eq:Bornpower}), and choosing a negative
exponent $r \sim - m$ for the weights, it appears that $J_1$ and $J_2$ become
the same in the $\ptt \to \infty$ limit. Hence, the effect of the weight is to
balance the leading and the second leading jet contributions.

\bibliographystyle{UTPstyle}
\bibliography{sijets}

\providecommand{\href}[2]{#2}\begingroup\raggedright\begin{thebibliography}{10}

\bibitem{Martin:1987vw}
A.~D. Martin, R.~G. Roberts, and W.~J. Stirling, {\it {Structure Function
  Analysis and psi, Jet, W, Z Production: Pinning Down the Gluon}},  {\em Phys.
  Rev.} {\bf D37} (1988) 1161.

\bibitem{Aversa:1988fv}
F.~Aversa, P.~Chiappetta, M.~Greco, and J.~P. Guillet, {\it {Higher Order
  Corrections to QCD Jets}},  {\em Phys. Lett.} {\bf B210} (1988) 225.

\bibitem{Ellis:1988hv}
S.~D. Ellis, Z.~Kunszt, and D.~E. Soper, {\it {The One Jet Inclusive
  Cross-section at Order $\alpha_s^3$: Gluons Only}},  {\em Phys. Rev. Lett.}
  {\bf 62} (1989) 726.

\bibitem{Currie:2016bfm}
J.~Currie, E.~W.~N. Glover, and J.~Pires, {\it {Next-to-Next-to Leading Order
  QCD Predictions for Single Jet Inclusive Production at the LHC}},  {\em Phys.
  Rev. Lett.} {\bf 118} (2017), no.~7 072002,
  [\href{http://xxx.lanl.gov/abs/1611.01460}{{\tt arXiv:1611.01460}}].

\bibitem{Currie:2018xkj}
J.~Currie, A.~Gehrmann-De~Ridder, T.~Gehrmann, E.~W.~N. Glover, A.~Huss, and
  J.~Pires, {\it {Infrared sensitivity of single jet inclusive production at
  hadron colliders}},  {\em JHEP} {\bf 10} (2018) 155,
  [\href{http://xxx.lanl.gov/abs/1807.03692}{{\tt arXiv:1807.03692}}].

\bibitem{Nagy:2003tz}
Z.~Nagy, {\it {Next-to-leading order calculation of three jet observables in
  hadron hadron collision}},  {\em Phys. Rev.} {\bf D68} (2003) 094002,
  [\href{http://xxx.lanl.gov/abs/hep-ph/0307268}{{\tt hep-ph/0307268}}].

\bibitem{Nagy:2001fj}
Z.~Nagy, {\it {Three jet cross-sections in hadron hadron collisions at
  next-to-leading order}},  {\em Phys. Rev. Lett.} {\bf 88} (2002) 122003,
  [\href{http://xxx.lanl.gov/abs/hep-ph/0110315}{{\tt hep-ph/0110315}}].

\bibitem{Ball:2017nwa}
{\bf NNPDF} Collaboration, R.~D. Ball et~al., {\it {Parton distributions from
  high-precision collider data}},  {\em Eur. Phys. J.} {\bf C77} (2017), no.~10
  663, [\href{http://xxx.lanl.gov/abs/1706.00428}{{\tt arXiv:1706.00428}}].

\bibitem{Buckley:2014ana}
A.~Buckley, J.~Ferrando, S.~Lloyd, K.~Nordström, B.~Page, M.~Rüfenacht,
  M.~Schönherr, and G.~Watt, {\it {LHAPDF6: parton density access in the LHC
  precision era}},  {\em Eur. Phys. J.} {\bf C75} (2015) 132,
  [\href{http://xxx.lanl.gov/abs/1412.7420}{{\tt arXiv:1412.7420}}].

\bibitem{Cacciari:2008gp}
M.~Cacciari, G.~P. Salam, and G.~Soyez, {\it {The anti-$k_t$ jet clustering
  algorithm}},  {\em JHEP} {\bf 04} (2008) 063,
  [\href{http://xxx.lanl.gov/abs/0802.1189}{{\tt arXiv:0802.1189}}].

\bibitem{Cacciari:2011ma}
M.~Cacciari, G.~P. Salam, and G.~Soyez, {\it {FastJet User Manual}},  {\em Eur.
  Phys. J.} {\bf C72} (2012) 1896,
  [\href{http://xxx.lanl.gov/abs/1111.6097}{{\tt arXiv:1111.6097}}].

\bibitem{Dasgupta:2016bnd}
M.~Dasgupta, F.~A. Dreyer, G.~P. Salam, and G.~Soyez, {\it {Inclusive jet
  spectrum for small-radius jets}},  {\em JHEP} {\bf 06} (2016) 057,
  [\href{http://xxx.lanl.gov/abs/1602.01110}{{\tt arXiv:1602.01110}}].

\bibitem{Cacciari:2003fi}
M.~Cacciari, S.~Frixione, M.~L. Mangano, P.~Nason, and G.~Ridolfi, {\it {The t
  anti-t cross-section at 1.8-TeV and 1.96-TeV: A Study of the systematics due
  to parton densities and scale dependence}},  {\em JHEP} {\bf 04} (2004) 068,
  [\href{http://xxx.lanl.gov/abs/hep-ph/0303085}{{\tt hep-ph/0303085}}].

\bibitem{Bellm:2019yyh}
J.~Bellm et~al., {\it {Jet cross sections at the LHC and the quest for higher
  precision}},  \href{http://xxx.lanl.gov/abs/1903.12563}{{\tt
  arXiv:1903.12563}}.

\bibitem{jetbook}
S.~Marzani, G.~Soyez, and M.~Spannowsky, {\it {Looking inside jets: an
  introduction to jet substructure and boosted-object phenomenology}},
  \href{http://xxx.lanl.gov/abs/1901.10342}{{\tt arXiv:1901.10342}}.

\end{thebibliography}\endgroup

\end{document}